\documentclass[12pt]{elsarticle}
\pdfoutput=1
\usepackage{graphicx,amsmath,color,microtype,fullpage,amsbsy,amsfonts,contour,bm}
\usepackage[]{algorithm}
\usepackage[skip=-18pt]{caption}
\usepackage{mathptmx}

\renewcommand\boldsymbol[1]{\bm{#1}} 

\usepackage{hyperref}
\definecolor{webgreen}{rgb}{0,.35,0}
\definecolor{webbrown}{rgb}{.6,0,0}
\definecolor{RoyalBlue}{rgb}{0,0,0.9}
\definecolor{mywhite}{rgb}{1.0,1.0,1.0}
\definecolor{mymagenta}{rgb}{0.93333,0.06667,0.53333}
\definecolor{mydblue}{rgb}{0.1176,0.1176,0.647}
\definecolor{mygray}{rgb}{0.68,0.68,0.68}
\definecolor{mygrayb}{rgb}{0.12,0.12,0.12}
\newcommand{\white}[1]{\textcolor{mywhite}{#1}}
\newcommand{\magenta}[1]{\textcolor{mymagenta}{#1}}
\newcommand{\dblue}[1]{\textcolor{mydblue}{#1}}

\hypersetup{
  colorlinks=true, linktocpage=true, pdfstartpage=3, pdfstartview=FitV,
  breaklinks=true, pdfpagemode=UseNone, pageanchor=true, pdfpagemode=UseOutlines,
  plainpages=false, bookmarksnumbered, bookmarksopen=true, bookmarksopenlevel=1,
  hypertexnames=true, pdfhighlight=/O, 
  urlcolor=webbrown, linkcolor=RoyalBlue, citecolor=webgreen,
  pdfauthor={Boris Valkov, Chris H. Rycroft, and Ken Kamrin},
  pdftitle={Eulerian method for fluid--structure interaction and submerged solid--solid contact problems},
  pdfsubject={},
  pdfkeywords={},
  pdfcreator={pdfLaTeX},
  pdfproducer={LaTeX with hyperref}%
}

\newcommand{\p}{\partial}
\renewcommand{\vec}[1]{\mathbf{#1}}
\newcommand{\ten}[1]{\mathbf{#1}}
\newcommand{\chib}{\boldsymbol\chi}
\newcommand{\bxi}{\boldsymbol\xi}

\newcommand{\bsig}{\boldsymbol\sigma}

\newcommand{\mattran}{\mathsf{T}}
\newcommand{\bv}{\vec{v}}
\newcommand{\bx}{\vec{x}}
\newcommand{\bX}{\vec{X}}
\newcommand{\bF}{\ten{F}}
\newcommand{\bB}{\ten{B}}
\newcommand{\cn}[1]{\makebox(0,0)[c]{#1}}
\newcommand{\Trans}{\Omega}
\newcommand{\Sep}{\gamma}
\newcommand{\krep}{k_\text{rep}}

\newcommand{\twid}{w_T}
\newcommand{\hspc}{h}
\newcommand{\nor}{\hat{\vec{n}}}
\newcommand{\ewid}{w_E}
\newcommand{\gind}[1]{\makebox(0,0)[r]{\textcolor{mygray}{#1}}}
\newcommand{\gindb}[1]{\makebox(0,0)[r]{\textcolor{mygrayb}{#1}}}

\DeclareMathOperator{\tr}{tr}

\definecolor{dkgreen}{rgb}{0,0.6,0}
\definecolor{gray}{rgb}{0.5,0.5,0.5}
\definecolor{mauve}{rgb}{0.58,0,0.82}
\definecolor{figred}{rgb}{0.6431,0.1765,0}
\definecolor{figblue}{rgb}{0.1922,0.1882,0.5333}

\begin{document}
\title{Eulerian method for fluid--structure interaction and submerged solid--solid contact problems}
\author[MIT1]{Boris Valkov}
\ead{bvalkov@mit.edu}
\author[Harvard,LBL]{Chris H. Rycroft}
\ead{chr@seas.harvard.edu}
\author[MIT1]{Ken Kamrin\corref{cor1}}
\ead{kkamrin@mit.edu}
\cortext[cor1]{Corresponding author.}
\address[MIT1]{Department of Mechanical Engineering, Massachusetts Institute of Technology, Cambridge, MA 02139, United States}
\address[Harvard]{School of Engineering and Applied Sciences, Harvard University, Cambridge, MA, 02138, United States}
\address[LBL]{Department of Mathematics, Lawrence Berkeley Laboratory, Berkeley, CA 94720, United States}

\begin{abstract}
We present a fully Eulerian, blurred-interface numerical method for fluid--structure interaction (FSI) with extension to the case of fluid-immersed solids interacting through contact. The method uses the Eulerian-frame Reference Map Technique (RMT) to represent the solid phase(s), permitting simulation of large-deformation constitutive behaviors. We demonstrate the method with multiple examples involving a compressible Navier--Stokes fluid coupled to a neo-Hookean solid. We verify the method's convergence. The algorithm is faster and more stable than previous methods based on RMT.  It is easily appended with a contact subroutine for multiple solids interacting within fluid, which we introduce and demonstrate with two examples.

\end{abstract}

\maketitle

\section{Introduction}

The challenges of simulating fluid--structure interaction (FSI) have been approached from many directions. One set of challenges stem from domain discretization: fluid problems on their own are amenable to solution on an Eulerian computational domain~\cite{chorin67,tannehill97,versteeg95,sethian99,hirt74} and solid deformation is natural to compute within a Lagrangian framework~\cite{zienkiewicz67,sulsky94,hoover06,belytschko00}. To couple solid and fluid phases in one setting, several approaches have been proposed to mitigate this separation in methodology. One approach is to treat the solid with a standard Lagrangian finite-element framework and to apply an Arbitrary-Lagrange Eulerian method in the fluid~\cite{bathe07,wang08,rugonyi01}, which remeshes the fluid domain in order to remedy issues of excessive mesh deformation. Other approaches include the family of immersed methods~\cite{bathe07,wang08,peskin02}, which keep an ambient stationary Eulerian grid throughout, on which fluid flow is solved, as well as a moving collection of interacting material points representing the solid structure. Here discretized delta functions are used to pass information between the grid and the nodes.

Some advantages of a fully Eulerian method---fluid and solid both computed on an Eulerian grid---can be directly seen. Since all phases are solved by sweeping through a single fixed mesh, there are computation time advantages. Topological changes are easy to manage on a fixed grid using level sets to track interfaces~\cite{sethian99,osher88}. Furthermore, multiscale and multiphysics coupling can also have advantages when done an Eulerian grid. One specific example, which will be discussed in more detail later, is the case of multiple solids making contact immersed in a fluid. Finding contact between two solid phases on an Eulerian grid can be achieved using grid-wise distance functions, or simply identifying grid points which become occupied by multiple solid phases during a trial step.

To achieve these goals one must pose an Eulerian scheme capable of solving finite-deformation solid problems. The recently proposed Reference Map Technique (RMT) is such an Eulerian framework, based on tracking the \emph{reference map} field~\cite{kamrin09, kamrinjmps}. Other approaches for solid deformation on a fixed Eulerian grid include hypoelastic implementations~\cite{udaykumar03,rycroft12} and methods that directly evolve the deformation gradient tensor field~\cite{plohr88,trangenstein91,liu01}. In a previous paper~\cite{kamrinjmps}, the RMT demonstrated the capability of accurately solving hyperelastic solid deformation problems on a fixed mesh---including shock propagation problems and problems with varied boundary and initial conditions---up to second-order accuracy in space and time. It also provided the first demonstrations of using the method to solve fully-coupled problems of FSI. There, the FSI method hinged on a \emph{sharp-interface representation}, extending on that of the Ghost Fluid Method for fluid--fluid interaction~\cite{fedkiw99}. Sharp methods make a distinct separation between each phase down to the sub-grid level. In the current work, our efforts exploit a \emph{blurred interface}, a simpler and computationally faster implementation, involving fewer numerical extrapolations. A blurred interface method uses a thin \textit{transition zone} where one phase converts into the other. As the grid size decreases, the corresponding transition zone decreases, and results approach that of a sharp interface method. For its use coupled to RMT for an FSI algorithm, the blurred interface method has certain advantages, which we will present in this work, and which extend the versatility of the FSI method to new applications.


To satisfy sub-grid jump conditions, sharp interface methods require a large number of grid-wise extrapolations of kinematic and stress fields across the interface. These produce the ``ghost values'' for each phase, which represent an extension of each field into the region occupied by the other phase(s). The validity of a sharp interface method is limited by the quality of continued and progressive extrapolation. When used in the FSI method of previous work~\cite{kamrinjmps}, accrued extrapolation error can have a destabilizing effect---shots of pressure along the interface may erroneously appear when the interface crosses through grid cell boundaries, in response to the sampling of extrapolated values when new points enter the solid domain. Adding significant solid dissipation or surface tension can penalize these artifacts, but this can alter the physicality of the simulation. We have found that errors of this type often produce routine-ending numerical instabilities, posing a serious implementational issue. By contrast, in fluid--fluid interaction methods, this effect is less important due to the natural viscous damping within all phases.

By switching the interfacial treatment to a blurred method we present a compromise wherein we move away from the sub-grid interfacial representation of a sharp technique in order to achieve a faster, simpler, and more stable method. Existing fluid--fluid blurred techniques~\cite{chern07,liu00} do not require any extrapolation. In the case of a fluid--solid problem as shown herein, using only one extrapolation---that of the reference map (described in more detail in section \ref{sec:rmap})---we are able to simulate FSI with a hyperelastic solid body on a single fixed grid using a single velocity field for the whole computational domain. In addition, jump conditions do not need to be explicitly applied but are implicitly met inside the transition zone. Moreover, as discussed in Sec.~\ref{FSSI}, the fact that the method uses one universal velocity field valid for all phases permits a straightforward approach for simulating multi-body contact of solids submerged in a fluid, a new capability for multi-phase RMT simulation.

This paper proceeds as follows. We first review the fundamental relations of each phase. The RMT methodology is then described, and we present the selected stencils for our finite-differences. Next, we define the transition zone and how the fields are appropriately mixed within its band of influence.  We then discuss a number of FSI test results and the convergence of the method. An important ingredient in our scheme is a new spatial extrapolation procedure that enforces certain physical and smoothness requirements in the field extrapolation. This is presented and shown to remedy deleterious extrapolation artifacts near the interface that can occur otherwise. We then describe how the routine can be appended with a solid--solid contact subroutine by taking advantage of the signed-distance measures inherent within the existing level set fields used in distinguishing the interfaces. Two examples are provided of submerged solid--solid contact.
  
\section{Theory}

\subsection{Eulerian-frame solid and fluid formulations}
\label{sec:rmap}
We first provide a brief review of the Eulerian formulation for solid simulation on which the RMT is based; details can be found in~\cite{kamrinjmps}. The motion function $\chib$ is defined as the time-dependent map from points $\bX$ in the reference configuration to their current position $\vec{x}$ in the deformed configuration; {\it i.e.} $\vec{x}=\chib(\vec{X},t)$. We define the spatial velocity field, $\bv=\bv(\bx,t)$, the material density, $\rho=\rho(\bx,t)$, and Cauchy stress, $\bsig=\bsig(\bx,t)$. In Eulerian-frame the conservation of mass and momentum in strong form (when deformations remain smooth) are
\begin{equation}
\rho_{t} = - \bv \cdot \nabla \rho - \rho\ \nabla \cdot \bv
\label{eq:mass2}
\end{equation}
and
\begin{equation}
  \bv_{t} = - (\bv \cdot \nabla) \bv - \frac{\nabla \cdot \bsig + \rho \mathbf{g}}{\rho},
\label{eq:momentum2}
\end{equation}
respectively, where $\vec{g}$ is the acceleration due to gravity, $\nabla$ is the gradient operator in the deformed space, and $\nabla\cdot$ is the spatial divergence operator.
The motion function permits us to define the deformation gradient as
\begin{equation}
  \bF(\textbf{X},t)=\frac{\partial\chib(\bX,t)}{\partial\bX}.
\end{equation}
We define the reference map $\bxi(\vec{x},t)$, an Eulerian field, as the inverse motion, so that 
\begin{equation}
\bX=\bxi(\bx,t)=\chib^{-1}(\bx,t).
\end{equation}
Because the reference map indicates the original location of a particle, the reference map for a tracer particle never changes, giving us an advective evolution law,
\begin{equation}
{\bxi_{t}+\bv\cdot\nabla\bxi=\boldsymbol{0}}.
\label{eq:rmap_update_cont}
\end{equation}
In the case where the initial configuration is undeformed ({\it i.e.} no
pre-strain) the initial condition for the reference map is
\begin{equation}
\bxi(\bx,t=0)=\bx=\bX.
\end{equation}
Using the chain rule it can then be shown that the deformation gradient \textbf{F} is
\begin{equation}
\bF(\bxi(\bx,t),t)=(\nabla\bxi(\bx,t))^{-1}
\label{eq:deformation_gradient2}
\end{equation}
and as such we can express the density in terms of the reference map and the
original density, $\rho_0$, by
\begin{equation}\label{solid_density}
  \rho=\frac{\rho_0}{\det \vec{F}} = \rho_0\det \nabla\bxi.
\end{equation}

Consider a solid body. Because the deformation gradient tensor is describable in terms of the reference map, we are thus capable of modeling the constitutive response of large-deformation, thermodynamically compatible solid laws in Eulerian frame. For example, given the reference map field corresponding to the deformation of an isotropic hyperelastic material with strain-energy density per reference volume $\psi_R$, the stress is given by
\begin{equation}\label{hyperstress}
\bsig=2(\det \bF)^{-1}\left.\frac{\partial\hat{\psi}_R(\bB)}{\partial\bB}\bB\right|_{\bF=(\nabla\bxi)^{-1}, \ \bB\equiv(\nabla\bxi)^{-1}(\nabla\bxi)^{-\mattran}},
\end{equation}
where $\bB$ is the left Cauchy--Green tensor and the $\ \hat{}\ $ indicates a constitutive function.
%
%
The specific model applied in this work will be a compressible neo-Hookean elastic solid model,
where the strain-energy density per reference volume is
\begin{equation}
\psi_R = \frac{G}{2}(J^{-\frac{2}{3}}\tr \bB -3)+\frac{\kappa}{2}(J-1)^{2}
\label{eq:neo-hookean2}
\end{equation}
and the Cauchy stress is
\begin{equation}
\bsig= GJ^{-5/3}\bB'+\kappa(J-1)\boldsymbol{1},
\label{eq:cauchy_stress_tensor2}
\end{equation}
where $J=\det \bF$ and a prime indicates the deviatoric part of a tensor. Equations \ref{eq:momentum2}, \ref{eq:rmap_update_cont}, \ref{eq:deformation_gradient2}, \ref{solid_density}, and \ref{hyperstress} are a closed Eulerian system for computing solid deformations. As shown in previous work~\cite{kamrinjmps}, this system of equations can be recast in conservative form and implemented numerically in a discrete conservation scheme should non-smooth solutions be expected.

Considering a fluid, the same equations of mass and momentum balance are valid but the constitutive law for the stress is different and no reference map is needed. In this work, we choose a weakly compressible viscous flow relation for fluid stress,
\begin{equation}
\bsig= \eta \frac{\nabla\bv+ (\nabla\bv)^\mattran}{2}-\lambda \left(\frac{\rho}{\rho_0}-1\right)\textbf{1}
\label{eq:fluid_stress_tensor}
\end{equation}
where $\eta$ is the viscosity and $\lambda$ is the compressibility modulus.

In the continuum limit, Eqs.~\ref{eq:mass2} and \ref{solid_density} give identical results. For later purposes of accurate discretization, it is preferable to implement Eq.~\ref{eq:mass2} when computing fluid density, since Eq.~\ref{solid_density} would require keeping a reference map field in a fluid, which would quickly lose accuracy due to the levels of distortion and mixing in the fluid. However, Eq.~\ref{solid_density} is preferable in a solid as it ensures consistency between the deformation and the density. Since we will soon consider coupled fluid--solid problems, it is helpful to distinguish between stresses and densities obtained from the different formulae. The solid stress and density have an ``s'' superscript, so that
\begin{equation}
\bsig^s = GJ^{-5/3}\bB'+\kappa(J-1)\boldsymbol{1}, \qquad \rho^s =\rho_0^s \det \nabla\bxi, \label{solid_sd}
\end{equation}
and the fluid stress and density have an ``f'' superscript, so that
\begin{equation}
\bsig^f = \eta \frac{\nabla\bv+ (\nabla\bv)^\mattran}{2}-\lambda \left(\frac{\rho^f}{\rho_0^f}-1\right)\textbf{1}, \qquad
\rho^f_{t} = - \bv \cdot \nabla \rho^f - \rho^f \nabla \cdot \bv.\label{fluid_sd}
\end{equation}

\setlength{\unitlength}{0.9bp}
\begin{figure}
  \begin{center}
    \begin{picture}(475,175)(0,0)
      \put(0,0){\includegraphics[scale=0.9]{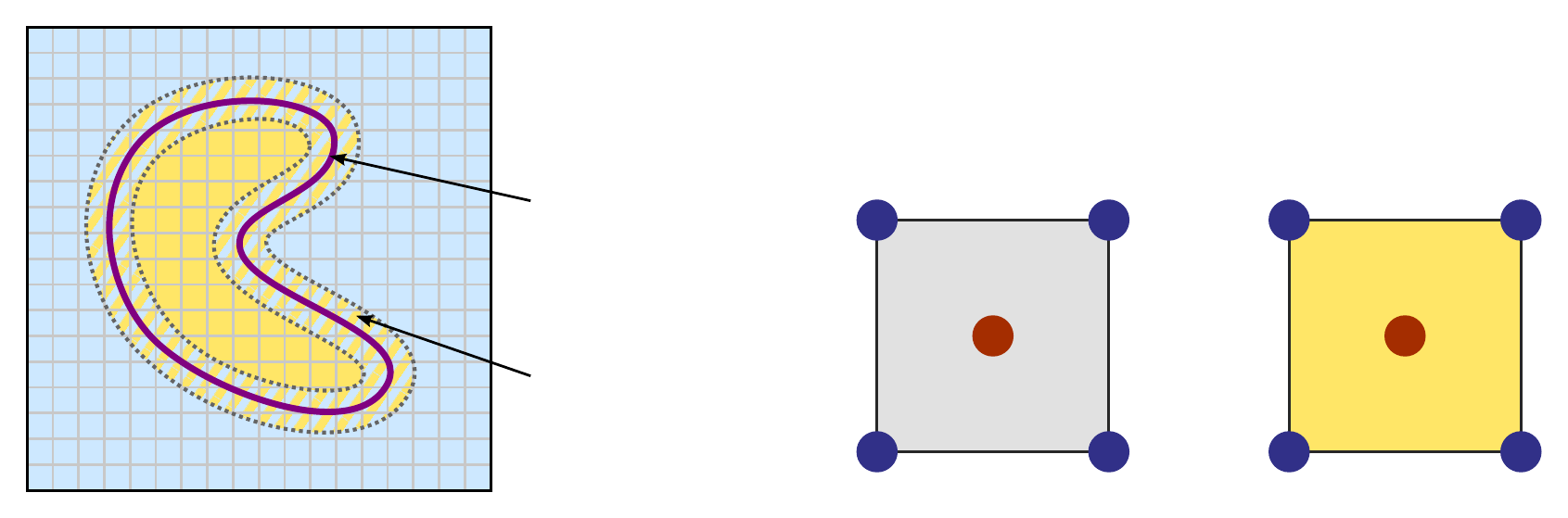}}
      \footnotesize
      \put(14,14){\bf{Fluid}}
      \put(57,107){\bf{Solid}}
      \put(188,99){\cn{Interface}}
      \put(188,86){\cn{$\phi(\bx,t)=0$}}
      \put(189,45){\cn{Transition}}
      \put(189,34){\cn{zone}}
      \put(189,21){\cn{$|\phi(\bx,t)|<\twid$}}
      \put(309,141){\cn{Globally defined}}
      \put(309,128){\cn{fields}}
      \put(438,141){\cn{Additional solid}}
      \put(438,128){\cn{fields}}
      \scriptsize
      \put(256,7){$i,j$}
      \put(317,7){$i+1,j$}
      \put(244,81){$i,j+1$}
      \put(304,81){$i+1,j+1$}
      \put(288,42){$[i,j]$}
      \put(385,7){$i,j$}
      \put(446,7){$i+1,j$}
      \put(373,81){$i,j+1$}
      \put(433,81){$i+1,j+1$}
      \put(417,42){$[i,j]$}
      \small
      \put(275,102){\textcolor{figblue}{$\bv,\rho^f,\rho,\phi$}}
      \put(312,64){\textcolor{figred}{$\bsig^f$}}
      \put(404,102){\textcolor{figblue}{$\bxi$}}
      \put(441,64){\textcolor{figred}{$\bsig^s,\rho^s$}}
    \end{picture}
  \end{center}
  \vspace{7mm}
  \caption{(Left) Overview of the simulation in which fluid and solid phases
  are simulated on a fixed Eulerian grid. The phases are determined by using a
  level set field $\phi(\bx,t)$, such that $\phi<0$ in the solid, $\phi>0$ in
  the fluid, and $\phi=0$ at the interface. In the numerical method, the
  constitive response between fluid and solid is blurred across a small
  transition zone, $|\phi|<\twid$. (Right) Arrangement of the simulation
  fields on each grid cell. The level set function $\phi$, fluid density
  $\rho^f$, fluid stress $\bsig^f$, combined density $\rho$, and velocity $\bv$
  are stored globally on all grid cells. In addition, the reference map field
  $\bxi$, solid stress $\bsig^s$, and $\rho^s$ are stored in the solid and slightly beyond (see text). As shown in the figure, some fields are stored
  at cell corners (blue) indexed by $i,j$ and some are stored at cell centers
  (red) indexed by $[i,j]$.\label{fig:setup}}
\end{figure}
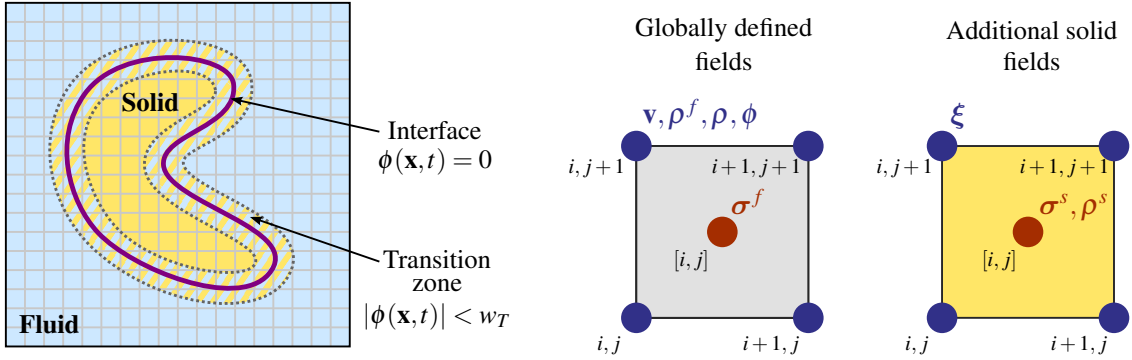

\subsection{The computational domain}
The computations are carried out on a two-dimensional $m \times n$ grid of
square cells with side length $h$, using plane strain conditions. The interface
between the fluid and solid phases is described using the level set
method~\cite{sethian99,osher03}, whereby an auxiliary Eulerian function
$\phi(\bx,t)$ is introduced such that the interface is given by the zero
contour, $\phi=0$. At the start of each simulation step, the level set
function is initialized to be the signed distance to the interface, using the
convention that $\phi<0$ inside the solid region and $\phi>0$ in the fluid
region. While there are a number of different numerical approaches for the
level set method, we make use of the specific implementation described by
Rycroft and Gibou~\cite{rycroft12}. The implementation contains a procedure
for reinitializing the level set function so that it is a signed distance
function to the zero contour, using a combination of the modified
Newton--Rapshon iteration of Chopp~\cite{chopp01} to update values of $\phi$
adjacent to the interface, and a second-order fast marching
method~\cite{sethian99} to update values further away. 

Because we develop a blurred interface technique, we introduce a transition
zone corresponding to the region where $|\phi|<\twid$, where $w_T$ is a
constant (Fig.~\ref{fig:setup}). As described in detail later, the material
response in the transition zone is modeled as a mixture of fluid and solid
phases. The region $0<\phi<\twid$ is more fluid than solid, and the region
$\twid<\phi<0$ is more solid than fluid. For a positive constant $\ewid$ such
that $\ewid>\twid$, we define the {\it extended solid domain} to be the region
$\phi<\ewid$, which corresponds to a region with any fraction of solid
response together with a thin band of points adjacent to the transition zone.
For computational efficiency our level set method implementation only stores
and updates the values of $\phi$ in a narrow band of grid points within a
distance $\pm\ewid$ of the interface.

Figure~\ref{fig:setup} shows the discretization of the simulation fields. Based
on considerations of the finite-difference stencils presented later, a
staggered field arrangement is used. Some fields are held at cell corners,
which are indexed using $i,j$ for $i=0,\ldots,m$ and $j=0,\ldots,n$. Some
fields are held at cell centers, which are indexed using $[i,j]$ for
$i=0,\ldots,m-1$ and $j=0,\ldots,n-1$. The level set function $\phi$, velocity
field $\bv$, fluid density \smash{$\rho^f$}, fluid stress \smash{$\bsig^f$},
and combined density $\rho$ are stored globally at cell corners in both the
fluid and solid domains. The fluid stress is stored at globally at cell
centers. In addition, in the extended solid domain, the reference map field
$\bxi$ is stored at cell corners, and the solid stress $\bsig^s$ and density
$\rho^s$ are stored at cell centers.



\section{Outline of the method}
\label{sec:outline_blurry}
We now describe the progression of the blurred interface method. Starting at a time step $n$ the algorithm builds the kinematic fields for the next step $n+1$. The stress fields are calculated at every time step as a mid-step calculation, but do not need to be. The algorithm below summarizes the major steps of the routine. Details are provided thereafter.
\par\vspace{\baselineskip}
\begin{algorithm}[H]
\textit{Given:} {$\bv^{n}$, $\bxi^{n}$, $\rho^{fn}$, and $\phi^n$}\\
\textit{Compute:} {$\bv^{n+1}$, $\bxi^{n+1}$, $\rho^{f(n+1)}$, and $\phi^{n+1}$}\\
\begin{enumerate}
 
 \item Compute the reference map gradient (Eq.~\ref{eq:grad_rmap_fd}) and use it to calculate the solid stress $\bsig^{sn}$ in the extended solid domain (Sec.~\ref{stencils}).\;
 
 \item Compute the velocity gradient and use it with $\rho^{fn}$ to compute the fluid stress $\bsig^{fn}$ in the entire domain (Eq.~\ref{eq:fluid_stress_tensor}).
  
 \item Apply mixing rule to fluid/solid stress and fluid/solid density (Sec.~\ref{sec:mixing}) to obtain the actual stress field $\bsig^n$ and density $\rho^n$ in the entire domain. \;
 
 \item Calculate the update to reference map (Eq.~\ref{eq:rmap_update}) in the region $\phi^n<0$.\;
 
 \item Move the level set field to $\phi^{n+1}$ (Sec.~\ref{extrap_algorithm}).\;
 
 \item Calculate the fluid density update (Eq.~\ref{eq:density_update}) on the entire domain.\;
 
 \item Calculate the update to the velocity field (Eq.~\ref{eq:velocity_update}).\; 
 
 \item Apply the previously computed updates to obtain $\bxi^{n+1}$, $ \vec{v}^{n+1}$, and $\rho^{f(n+1)}$.\;
 
 \item Set boundary conditions and kinematic constraints.\;
 
 \item Extrapolate reference map into the portion of the extended solid domain where $\phi>0$ to obtain $\bxi^{n+1}$ in that region (Sec.~\ref{extrap_algorithm}).\; 
 \end{enumerate}
 \caption{Outline of the blurred interface method for FSI.\label{main_alg}}
 \end{algorithm}
\par\vspace{\baselineskip}

\subsection{Finite-difference stencils}\label{stencils}
We employ a tight stencil to calculate the divergence of stress and the (constitutive input) gradients of $\bxi$ and $\bv$ as in Fig.~\ref{fig:rmap_grad2}, with more details on our stencil selection to be discussed later. The discretized gradient of the reference map is given by
\begin{equation}
\left.\frac{\partial\bxi}{\partial x}\right|^n_{[i,j]} = \frac{\bxi^n_{i+1,j} - \bxi^n_{i,j}}{\hspc}, \qquad
\left.\frac{\partial\bxi}{\partial y}\right|^n_{[i,j]} = \frac{\bxi^n_{i,j+1} - \bxi^n_{i,j}}{\hspc}.
\label{eq:grad_rmap_fd}
\end{equation}
The discretization of $\nabla \vec{v}$ is identical to Eq.~\ref{eq:grad_rmap_fd}. The discretization of divergence of stress, $\nabla\cdot\bsig$, uses the following stencils for the stress derivatives:
\begin{equation}
\left.\frac{\partial\bsig}{\partial x}\right|^n_{i,j} = \frac{\bsig^n_{[i,j]} - \bsig^n_{[i-1,j]}}{\hspc},\qquad
\left.\frac{\partial\bsig}{\partial y}\right|^n_{i,j} = \frac{\bsig^n_{[i,j]} - \bsig^n_{[i,j-1]}}{\hspc}.
\label{eq:div_stress_fd}
\end{equation}
In the extended solid domain, $\bsig^s$ and $\rho^s$ are computed directly from the above calculated reference map gradient as applied to Eq.~\ref{solid_sd}. For example,
\begin{equation}
  \vec{B}^n_{[i,j]}=(\nabla\bxi^n_{[i,j]})^{-1}(\nabla\bxi^n_{[i,j]})^{-\mattran}
\end{equation}
and \smash{$J^n_{[i,j]}=\det \nabla\bxi^n_{[i,j]}$} would be applied to create \smash{$\bsig^{sn}_{[i,j]}$} and \smash{$\rho^{sn}_{[i,j]}$}. When put together, the divergence of the solid stress at $i,j$ is calculated from nearby values of $\bxi$ at $i,j$, at $i-1,j$, at $i+1,j$, at $i,j-1$, at $i,j+1$, at $i-1,j+1$, and at $i+1,j-1$.

\setlength{\unitlength}{0.9bp}
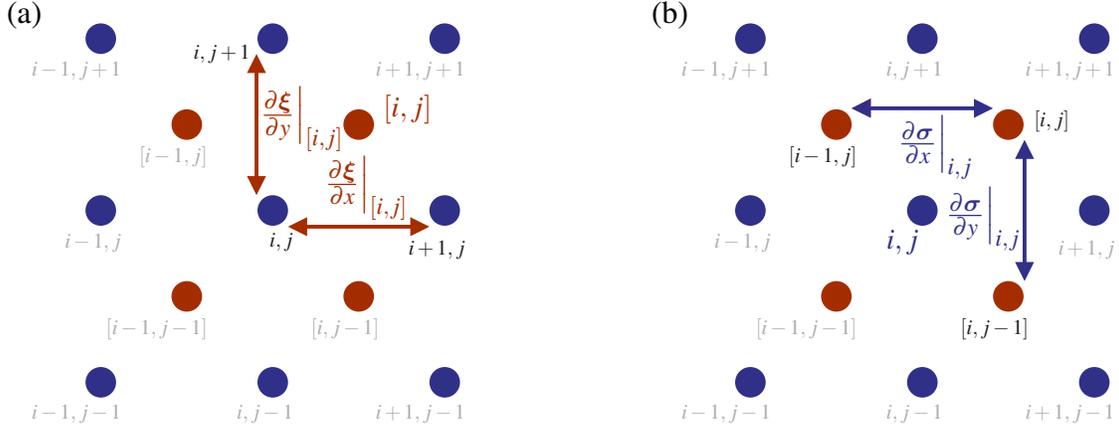
\begin{figure}
  \begin{center}
    \begin{picture}(460,195)(0,0)
      \put(25,30){\includegraphics[scale=0.9]{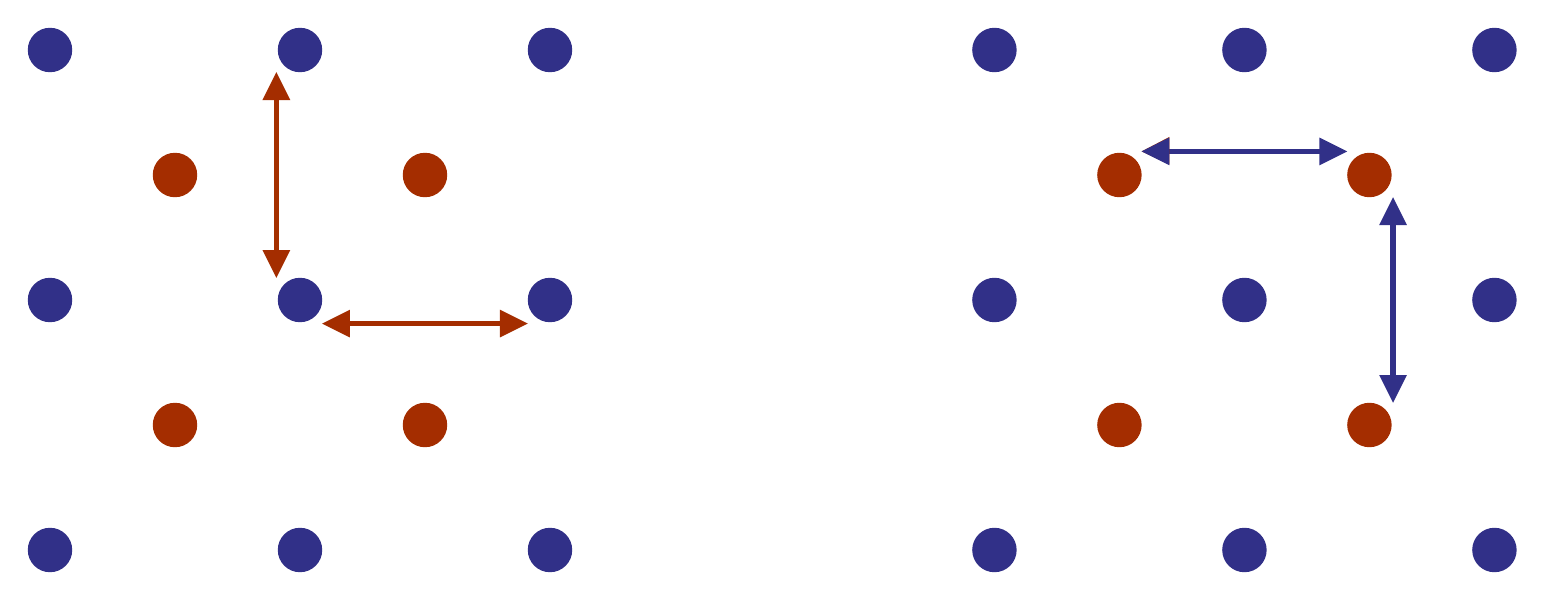}}
      \put(0,195){(a)}
      \put(270,195){(b)}
      \scriptsize
      \put(48,31){\gind{$i-1,j-1$}}
      \put(48,103){\gind{$i-1,j$}}
      \put(48,175){\gind{$i-1,j+1$}}
      \put(84,67){\gind{$[i-1,j-1]$}}
      \put(84,139){\gind{$[i-1,j]$}}
      \put(120,31){\gind{$i,j-1$}}
      \put(120,103){\gindb{$i,j$}}
      \put(102,182){\gindb{$i,j+1$}}
      \put(156,67){\gind{$[i,j-1]$}}
      \put(192,31){\gind{$i+1,j-1$}}
      \put(192,100){\gindb{$i+1,j$}}
      \put(192,175){\gind{$i+1,j+1$}}
      \small
      \put(178,159){\gind{\textcolor{figred}{$[i,j]$}}}
      \normalsize
      \put(132,124){\textcolor{figred}{$\left.\frac{\p \bxi}{\p x}\right|_{[i,j]}$}}
      \put(105,153){\textcolor{figred}{$\left.\frac{\p \bxi}{\p y}\right|_{[i,j]}$}}
      \scriptsize
      \put(320,31){\gind{$i-1,j-1$}}
      \put(320,103){\gind{$i-1,j$}}
      \put(320,175){\gind{$i-1,j+1$}}
      \put(356,67){\gind{$[i-1,j-1]$}}
      \put(356,139){\gindb{$[i-1,j]$}}
      \put(392,31){\gind{$i,j-1$}}
      \put(392,175){\gind{$i,j+1$}}
      \put(428,67){\gindb{$[i,j-1]$}}
      \put(445,154){\gindb{$[i,j]$}}
      \put(464,31){\gind{$i+1,j-1$}}
      \put(464,100){\gind{$i+1,j$}}
      \put(464,175){\gind{$i+1,j+1$}}
      \small
      \put(382,103){\makebox(0,0)[r]{\textcolor{figblue}{$i,j$}}}
      \normalsize
      \put(392,111){\textcolor{figblue}{$\left.\frac{\p \bsig}{\p y}\right|_{i,j}$}}
      \put(372,141){\textcolor{figblue}{$\left.\frac{\p \bsig}{\p x}\right|_{i,j}$}}
    \end{picture}
  \end{center}
  \vspace{2mm}
  \caption{Stencils for (a) the reference map gradient at $[i,j]$ per
  Eq.~\ref{eq:grad_rmap_fd} and (b) the divergence of stress at $i,j$ per
  Eq.~\ref{eq:velocity_update}. The cell-cornered grid is shown as blue circles
  and the cell-centered grid is shown as red circles. The indexes of grid
  points that are used in the stencils are shown in black, while those that are
  not used are shown in gray.\label{fig:rmap_grad2}}
\end{figure}

In computing fluid stress, the non-viscous fluid pressure is calculated as\footnote{This definition is such that, under Eq.~\ref{eq:div_stress_fd}, the stencil for divergence of $p^{fn}\textbf{1}$ at $i,j$ is given by one-sided differences at $i,j$. }
\begin{equation}
p^{fn}_{[i,j]} = \lambda \left(\frac{\rho^{fn}_{i,j}}{\rho^f_0}-1\right).
\label{eq:fluid_pressure}
\end{equation}
The fluid stress tensor is then given by
\begin{equation}
  \bsig^{fn}_{[i,j]} = \eta \frac{\nabla\bv^n_{[i,j]} + (\nabla\bv)^{n\mattran}_{[i,j]}}{2}-p^{fn}_{[i,j]}\textbf{1},
\label{eq:fluid_stress_tensor2}
\end{equation}
where the discretization of $\nabla \bv$ uses the same stencil as in Eq.~\ref{eq:grad_rmap_fd}. 

Now we discuss the stencil for the fields that are updated each step. The reference map is updated using Eq.~\ref{eq:rmap_update_cont}, which is discretized to
\begin{equation}
\bxi^{n+1}_{i,j}=\bxi^n_{i,j}-\Delta t \left[( \bv \cdot \nabla)\bxi)\right]^n_{i,j},
\label{eq:rmap_update}
\end{equation}
where the advection term $(\bv \cdot \nabla) \bxi$ is calculated using a
second-order, upwinded ENO discretization~\cite{shu88,rycroft12} for stability.
The update to velocity is given by
\begin{equation}
\bv^{n+1}_{i,j} = \bv^{n}_{i,j}+\Delta t\left( \left[(\bv\cdot\nabla)\bv\right]^n_{i,j} +\frac{1}{\rho_{i,j}}\left((\nabla\cdot\bsig^n)_{i,j}+\textbf{f}_{i,j}+{\rho_{i,j}}\vec{g} \right)\right),
\label{eq:velocity_update}
\end{equation}
where the divergence of stress is discretized as in Eq.~\ref{eq:div_stress_fd} and the advection term uses the ENO discretization.  Unless otherwise stated, we choose the gravity $\vec{g}$ to be zero in the results that follow. The above equation permits an additional body force $\vec{f}_{i,j}$
which we call upon in Sec.~\ref{FSSI}. A global damping term $\eta_a (\nabla^2 v)_{i,j}$ with artificial viscosity $\eta_{a}$ is also permitted as part of $\vec{f}_{i,j}$, but we ensure $\eta_{a}$ is small in comparison to $\eta$ so as to minimally affect the physics. A standard second-order, five-point stencil is used in computing this Laplacian. The fluid density is updated using
\begin{eqnarray}
\rho^{f (n+1)}_{i,j} &=& \rho^{fn}_{i,j} + \Delta t\left(-\left[(\bv\cdot\nabla)\rho^f\right]^n_{i,j}-\rho^{fn}_{i,j}\ (\nabla\cdot\bv)_{i,j}^n \right), \nonumber \\
&=& \rho^{fn}_{i,j} + \Delta t\left(-\left[(\bv\cdot\nabla)\rho^f\right]^n_{i,j}-\rho^{fn}_{i,j}\left(\frac{\bv_{i+1,j}^n-\bv_{i,j}^n}{h} + \frac{\bv_{i,j+1}^n -\bv_{i,j}^n}{h}\right) \right)
\label{eq:density_update}
\end{eqnarray}
where the advection term uses the ENO discretization, and the discretization of the fluid divergence is chosen in a similar manner to Eq.~\ref{eq:grad_rmap_fd}.

\subsection{Mixing quantities in the transition zone}
\label{sec:mixing}
We construct a transition field centered about the interface, which determines the fraction of solid-like behavior ({\it i.e.} density and stress) a point experiences, with the rest being attributed to fluid. This field defines how one phase cross-fades into the other through the mixing rules defined below. We base our transition field on the smoothed Heaviside function
\begin{equation}
  \label{eq:heavi_smooth}
  H_s(x) = \left\{
  \begin{array}{ll}
    0 & \qquad \text{if $x\le-\twid$,} \\
    \frac{1}{2}(1+\frac{x}{\twid}+\frac{1}{\pi}\sin\frac{\pi x}{\twid}) & \qquad \text{if $|x|<\twid$,} \\
    1 & \qquad \text{if $x\ge \twid$,}
  \end{array}
  \right.
\end{equation}
where $\twid$ is the transition zone width that was introduced previously. The choice of this function is not unique, although the above has the benefit of smoothly transitioning from exactly zero to exactly one over a finite interval, and it has been used in other blurred-interface methods~\cite{yu03,yu07}. We use the level set function as the input to the $H_s$, since it measures the distance from the interface. We define cell-cornered and cell-centered transition fields as
\begin{equation}
  \Trans_{[i,j]} = H_s(\phi_{[i,j]}), \qquad \Trans_{i,j} = H_s(\phi_{i,j}),
\label{eq:erf2}
\end{equation}
respectively, where \smash{$\phi_{[i,j]}$} is calculated by averaging the four
values on the grid cell corners. 

We now have a straightforward way of defining the mixture of quantities everywhere in our computational domain.  For instance, the stress components become a mixture of the stress obtained from the solid constitutive law and that of the fluid constitutive law weighted respectively by $\Trans$ and $1-\Trans$, so that
\begin{equation}
  \bsig_{[i,j]} = \Trans_{[i,j]} \bsig^{s}_{[i,j]} + (1-\Trans_{[i,j]}) \bsig^{f}_{[i,j]}.
\label{eq:s11mix}
\end{equation}
The global density field is defined similarly as
\begin{equation}
  \rho_{i,j} = \Trans_{i,j} \rho^{f}_{i,j} + (1-\Trans_{[i,j]})\frac{\rho^{s}_{[i,j]}+\rho^{s}_{[i+1,j]}+\rho^{s}_{[i,j+1]}+\rho^{s}_{[i+1,j+1]}}{4},
\label{eq:rhom}
\end{equation}
where the above uses the average of four values to interpolate $\rho^s$ onto the cell-cornered grid.

The above mixing procedure models the \emph{no slip} condition between fluid and solid. Other interfacial conditions are possible within this framework, though we have yet to perform extensive tests on them. For example, perfect slip may be achievable by applying a shear traction elimination step after the above mixing is applied, which modifies the stress in the transition zone by smoothly decreasing the shear traction resolved onto the level set contours such that shear traction vanishes at $\phi=0$.

\subsection{Reference map extrapolation}
In order to define $\bxi$ and consequently $\bsig^s$ in the portion of the
extended solid domain where $\phi>0$, great care must be taken. This zone
represents a region whose behavior is more fluid than solid, and, thus, is
liable to undergo extensive shearing and nonlinear deformation. This is
problematic because large nonlinear deformations reduce the accuracy of
computation of $\nabla\bxi$ necessary to calculate $\bsig^s$, and, moreover,
the solid stress can become unboundedly large as deformation builds up in this
zone, which negates the purpose of the gradual switch-over to a fluid-dominated
stress response in Eq.~\ref{eq:s11mix}. For these reasons we avoid advecting
the reference map per Eq.~\ref{eq:rmap_update} in this region and instead
obtain values for it through extrapolation from the nearby grid points where
$\phi<0$. This constitutes the only field extrapolation in our current method,
and it is recalculated every time step. Note that in the velocity update step,
only those values of stress within the transition zone ({\it i.e.} very close
to the $\phi=0$ contour) influence the outcome, which supports our usage of
near-field extrapolation as an appropriate remedy.  Our algorithm for
consistent extrapolation is left for a detailed discussion in
Sec.~\ref{extrap_algorithm}.

\section{Results}
The above comprises a general fluid--solid interaction routine for simulating solids with finite-strain constitutive relations coupled to a fluid phase that obeys the compressible Navier--Stokes equations. We now verify the method and its accuracy using several test geometries, many of which (those of Sec.~\ref{rod}) under the previous sharp-interface approach~\cite{kamrinjmps}, exhibited interfacial errors of the type previously described, even with the inclusion of surface tension and significant solid viscosity. Our current blurred-interface routine has also reduced the number of extrapolation steps significantly---twenty scalar fields are extrapolated per time-step in the sharp-interface approach, now only the two components of $\bxi$ are now extrapolated.  Unlike the sharp approach, we do not require a specific routine to ensure interfacial jump conditions as they are implicitly satisfied in the transition zone. These comments are not to suggest that the numerical interfacial errors of the sharp approach are irreparable in that framework; it is possible they could be alleviated in part through integration of our improved extrapolation algorithm of Sec.~\ref{extrap_algorithm} among other adjustments, but we reserve this topic for future work.

Throughout this paper, we make use of dimensionless simulation units. In all tests shown, unless otherwise stated, the following input material parameters are used. The initial fluid density is \smash{$\rho^f_0=1.0$}, the viscosity is $\eta=0.12$, and the compressibility is $\lambda=60.0$. The solid uses $G=10$ and $\kappa=50$. The artificial viscosity is $\eta_a=0.012$. These parameters match our previous work~\cite{kamrinjmps}, though it should be noted the previous work also included a significant separately added solid viscosity.
The predominant Courant--Friedrichs-Lewy (CFL) criterion is due to the fluid viscosity, which is proportional to $\hspc^2\rho^f/2\eta$. The other criteria that need to be considered are the one due to compressibility of the solid \smash{$\sim \hspc\sqrt{\rho^s/\kappa}$}, the one due to compressibility of the fluid \smash{$\sim \hspc\sqrt{\rho^f/\lambda}$}, and the convective CFL \smash{$\sim \hspc/\max |\vec{v}|$}. For all of our results, we use a timestep of $\Delta t=0.05h^2\rho^f/\eta$, which satisfies the dominant CFL condition for the fluid viscosity, and also satisfies the additional CFL conditions for all of the grids and parameters that are used. Throughout, we use a transition zone width of $\twid=3\hspc$. The simulations are written in C++ and use the OpenMP library to multithread many of the operations that scan over the entire grid of points.

For all of the simulations presented here, domain boundary conditions are applied by fixing the values of $\bv$ and $\rho^f$ on the edge of the cell-cornered grid, where $i=0$, $j=0$, $i=m$, or $j=n$. We consider two types of boundary conditions: Dirichlet conditions where the boundary field values are fixed, and free boundary conditions where the boundary field values are linearly extrapolated from the two adjacent layers of interior points. For example, to implement a free boundary condition on an arbitrary field $f_{i,j}$ at the top boundary, we set $f_{i,n} = 2f_{i,n-1}-f_{i,n-2}$ for $i=0,1,\ldots,m$.  Note that a free condition on the fluid velocity component tangent to a boundary gives the perfect-slip condition.

\subsection{Incoming fluid deforms anchored rubber rod}\label{rod}
The first example we consider involves a bulky, highly-deformable neo-Hookean rod. The rod is initially vertical and anchored at its top end. It is placed in a horizontal fluid flow, which causes it to deform significantly. The domain covers $-2\le x \le 2$ and $-2 \le y \le 2$ using a square grid of size $240\times240$. The bar initially covers the rectangle $-1.4<x<0.6$ and $|y|<1.1$ and has semi-circular end caps. The anchored region is a circle with center $(x,y)=(-1,1.1)$ and radius $r_a$. During the simulation the reference map in the anchored region is enforced to be constant and the velocity is enforced to be zero. We have simulated two different anchor sizes $r_a=0.25$ (Fig.~\ref{fig:stick_result_1}) and $r_a=0.15$ (Fig.~\ref{fig:stick_result_2}). The fluid inflow and outflow is controlled by applying a constant horizontal velocity of $(u,v)=(0.24,0)$ on the left and right sides of the domain window. Perfect slip boundary conditions are used on the top and bottom sides, where $v=0$, and $u$ and $\rho^f$ are free. In each simulation the fluid flow deforms the rod inducing large local stretches in the solid, with stretch ratios exceeding two in parts of the bar. In the case of the smaller anchor, the bar is less able to resist the incoming fluid flow and swivels out of the way to a greater extent as expected. The simulations each reach a steady state by $t\approx20$ where the rod remains in a static, bent configuration with steady fluid flow surrounding it. On an Apple MacBook Pro (Mid 2014) system with a quad-core 2.8~GHz Intel i7 processor, the simulation using $r_a=0.25$ takes 997~s using 216,090 timesteps using a single thread. If two, three, or four threads are used, the simulation take $773~s$, $705~s$, and $635~s$ respectively. Because this simulation only uses a small grid, the speedup from multithreading is only modest, since the overhead from creating threads is comparable the computational work done each timestep. However, for some of the larger simulation grids considered later, multithreading becomes significantly more advantageous.

We also used this configuration with $r_a=0.25$ to test the speed of the simulation method against the previous sharp interface method~\cite{kamrinjmps}. For a benchmark test simulating over the interval $0\le t \le 2.5$ with $3,200$ timesteps using a single thread, the sharp interface method took 94.01~s, while the new method took 53.97~s, corresponding to a speedup factor of 1.74. These two simulations were not perfectly comparable, since the memory organization and boundary implementation are slightly different. However, the results demonstrate that the new method is significantly faster, due to the simplification of the finite-difference stencils, a reduction in the number of simulation fields required, and fewer boundary extrapolations.

\setlength{\unitlength}{0.0125bp}
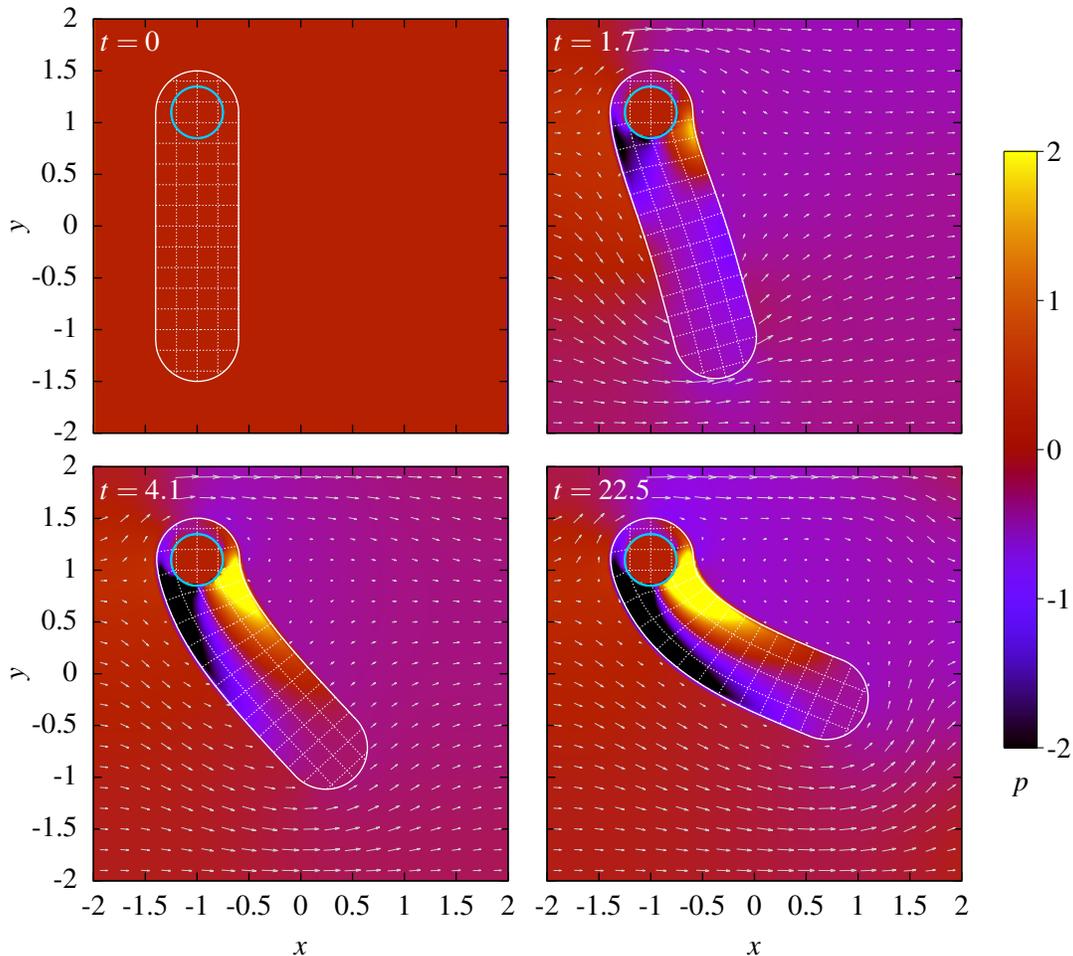
\begin{figure}
  \begin{center}
    \small
    \begin{picture}(33000,28600)(0,0)
      \put(-1500,13500){\parbox{15cm}{\include{stick1_p}}}
      \setlength{\unitlength}{0.0125bp}
      \put(30000,4700){ 
	\begin{picture}(2000,21100)(0,0)
	  \put(0,1600){\includegraphics[width=12.5pt,height=226pt]{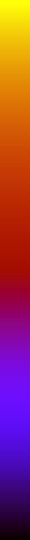}}
	  \put(0,1600){\line(0,1){18000}}
	  \put(0,1600){\line(1,0){1200}}
	  \put(1000,1600){\line(0,1){18000}}
	  \put(0,19600){\line(1,0){1200}}
	  \put(500,400){\makebox(0,0)[c]{$p$}}
	  \put(1000,6100){\line(1,0){200}}
	  \put(1000,10600){\line(1,0){200}}
	  \put(1000,15100){\line(1,0){200}}
	  \put(1300,1600){\makebox(0,0)[l]{-2}}
	  \put(1300,6100){\makebox(0,0)[l]{-1}}
	  \put(1300,10600){\makebox(0,0)[l]{0}}
	  \put(1300,15100){\makebox(0,0)[l]{1}}
	  \put(1300,19600){\makebox(0,0)[l]{2}}
	\end{picture}}
    \end{picture}
  \end{center}
  \vspace{6mm}
  \caption{Four snapshots of the in-plane pressure $p\equiv\frac{\sigma_{xx}+\sigma_{yy}}{2}$ (colors) and fluid velocity
  (arrows) for the simulation of an anchored flexible rod deformed to large
  strain by incoming fluid flow.  The solid white line shows the
  boundary of the rod, given by the zero contour of the level set function
  $\phi(\bx,t)$. The thin dashed white lines are the contours of the reference
  map $\bxi(\bx,t)$. The cyan circle of radius 0.25 shows the anchored region
  where the reference map is fixed and the velocity is
  zero.\label{fig:stick_result_1}}
\end{figure}

\begin{figure}
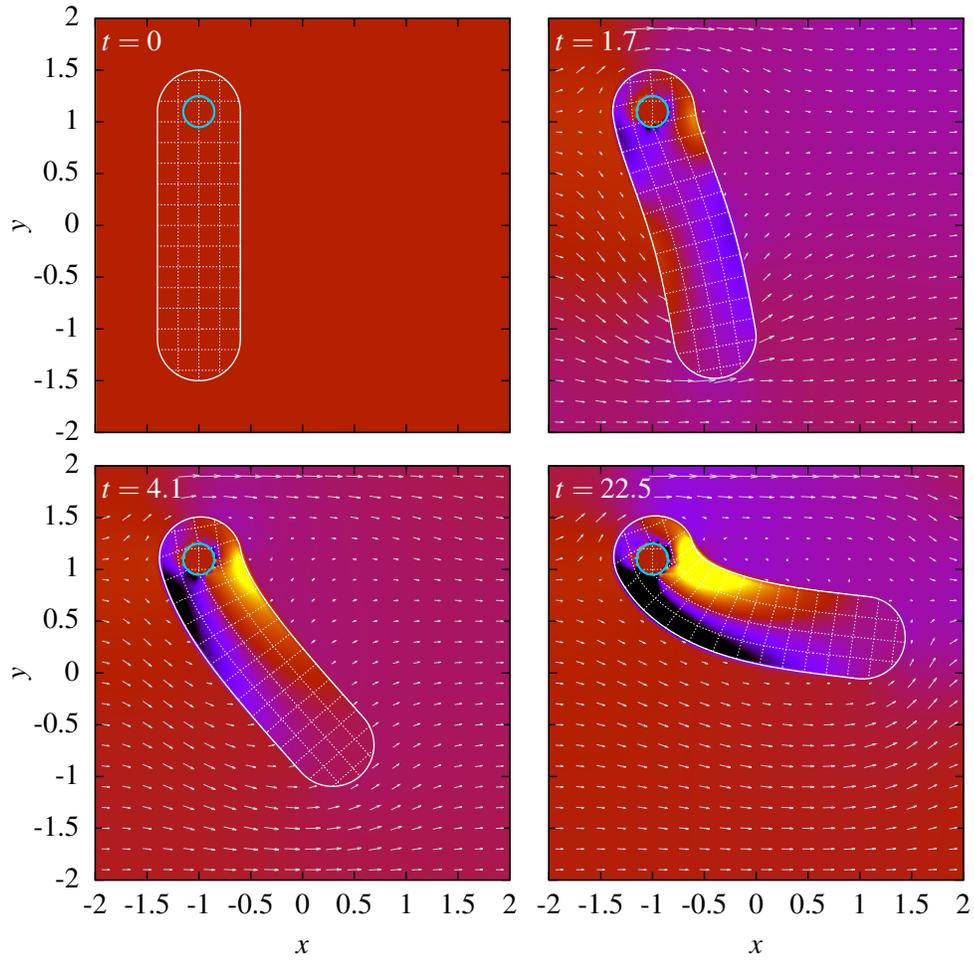

  \begin{center}
    \small
    \include{stick2_p}
  \end{center}
  \caption{Four snapshots of the in-plane pressure field (colors) and fluid velocity
  (arrows) for the simulation of an anchored flexible rod, using a smaller
  anchored region than in Fig.~\ref{fig:stick_result_1}. The solid white line
  shows the boundary of the rod, given by the zero contour of the level set
  function $\phi(\bx,t)$. The thin dashed white lines are the contours of the
  reference map $\bxi(\bx,t)$. The cyan circle of radius 0.15 shows the
  anchored region where the reference map is fixed and the velocity is
  zero.\label{fig:stick_result_2}}
\end{figure}

\begin{figure}
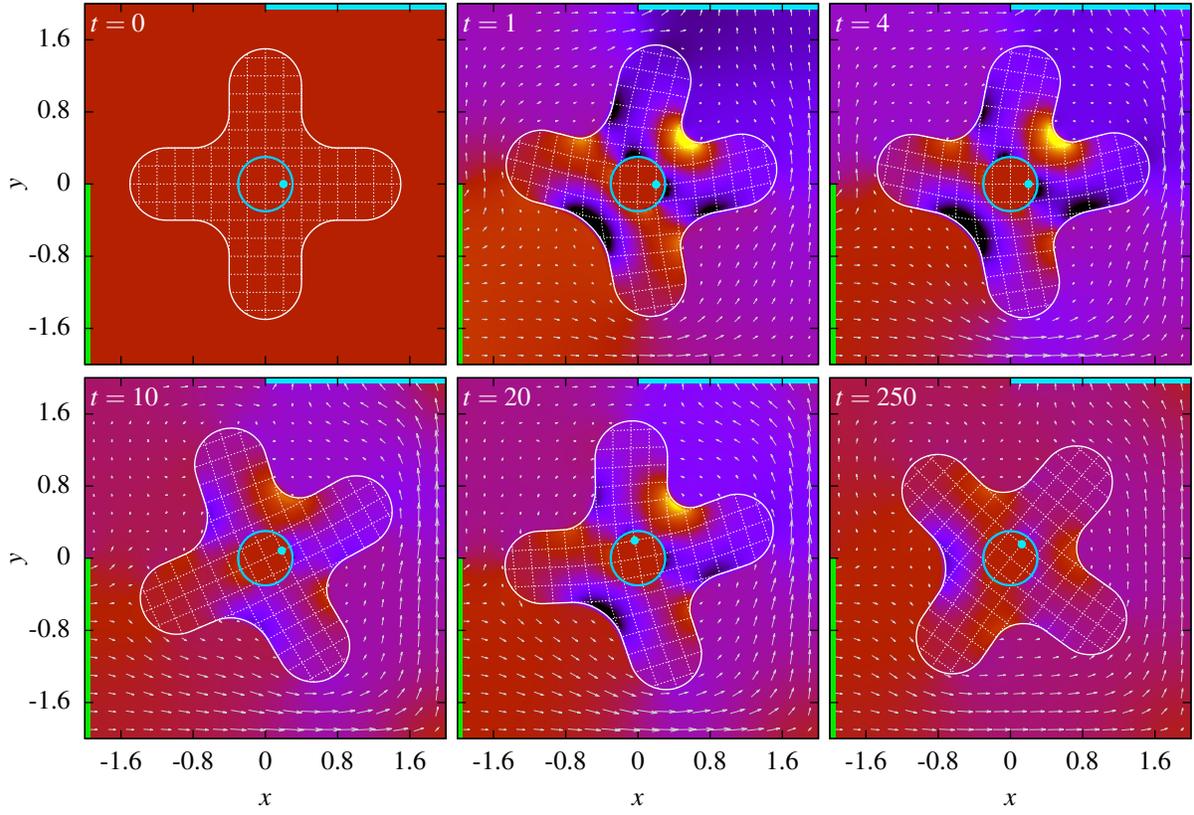

  \begin{center}
    \footnotesize
    \include{rotor_p}
  \end{center}
  \caption{Six snapshots of the in-plane pressure field (colors) and fluid velocity
  (arrows) for the simulation of a flexible rotor.  The green rectangle shows the
  region where fluid is added, and the cyan rectangle shows where fluid is
  removed. The solid white line shows the boundary of the solid, given by the
  zero contour of the level set function. The thin dashed white lines are the
  contours of the reference map $\bxi$. The cyan circle of radius 0.3 shows the
  pivot, and the small circular dot shows how the pivot has rotated. By
  $t=250$ the rotor has undergone roughly \smash{$5\frac{1}{2}$} complete
  rotations.\label{fig:rotor_result_1}}
\end{figure}

\begin{figure}
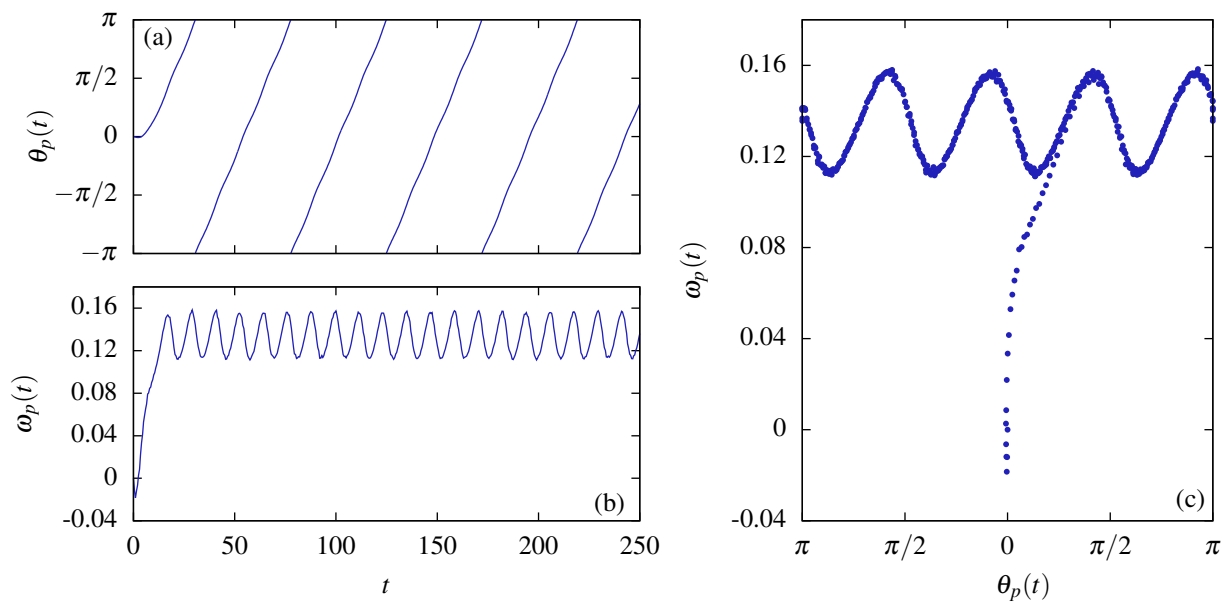

  \begin{center}
    \footnotesize
    \include{rotor_spin}
  \end{center}
  \vspace{2mm}
  \caption{Plot of (a) rotation angle $\theta_p$ and (b) angular velocity
  $\omega_p$ of the pivot as a function of time $t$ for the flexible rotor
  simulation shown in Fig.~\ref{fig:rotor_result_1}. (c) Plot of the
  relationship between $\theta_p$ and $\omega_p$. \label{fig:rotor_spin}}
\end{figure}

\subsection{Fluid deforms and spins a four-blade rotor}
\label{sub:rotor}
The second test case we consider involves a rotor of neo-Hookean material anchored around a pivot at its center. It is deformed and ultimately set into steady rotation by incoming fluid. Each blade of the rotor is comprised of a rectangle of length 1.1 with a semi-circular end cap. The join between each pair of blades is smoothed with a quarter-circle of radius 0.4. The fluid boundary condition consists of a prescribed fluid inflow of $(u,v)=(0.2,0)$ on the bottom half of the left edge of the computational domain, and an outflow of of $(u,v)=(0,0.2)$ in the top right half edge of the domain; at all other boundaries, perfect slip boundary conditions are used. The rotor and fluid have the same material parameters as the previous subsection. This particular geometry, when implemented under the previous sharp-interface method, displayed many erroneous pressure shots about the interface, as the particular shape involved has much curvature variation. As can be seen from Fig.~\ref{fig:rotor_result_1} this artifact is non-existent in the current scheme. By $t=14$ the rotor motion and fluid flow approach a steady repeating cycle, with each cycle corresponding to one quarter-turn of the rotor. By the final snapshot in Fig.~\ref{fig:rotor_result_1} at $t=250$, the rotor as undergone approximately $5\frac{1}{2}$ complete rotations without any interfacial perturbations. Figure~\ref{fig:rotor_spin} confirms that the rotor enters a steady repeating cycle, by showing the rotation angle $\theta_p$ and angular velocity $\omega_p$ of the pivot as a function of time. Figure~\ref{fig:rotor_spin}(c) shows a plot of $\theta_p$ against $\omega_p$, where after an initial transient period as the rotor begins to turn, a steady relationship between $\omega_p$ and $\theta_p$ develops, with the angular velocity being slightly higher during periods when a blade of the rotor is directly in front of the region of fluid inflow.

\begin{figure}
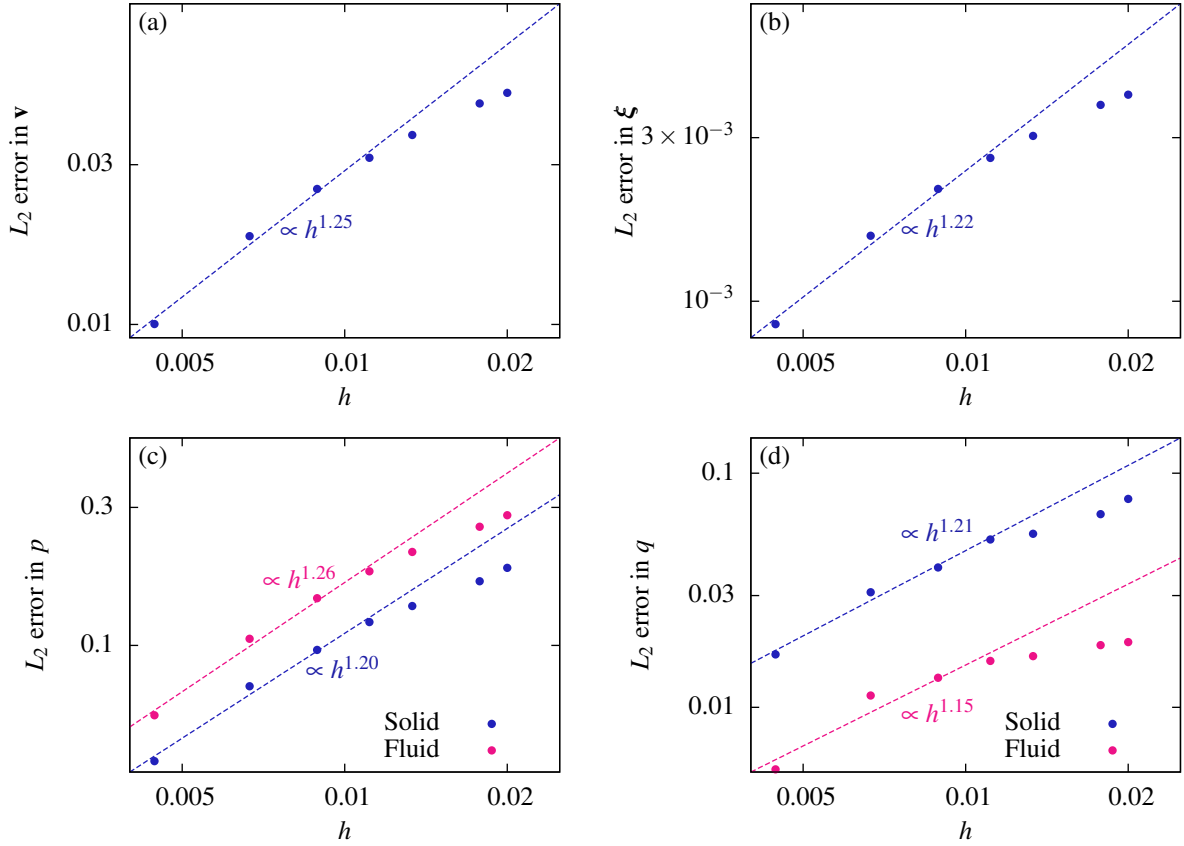

  \begin{center}
    \footnotesize
    \include{l2error}
  \end{center}
  \caption{Log--log plots of the $L_2$ error for the flexible rotor simulation
  at $t=0.4$ as a function of grid spacing $h$ for (a) velocity $\bv$, (b) the
  reference map $\bxi$, (c) $p$ as calculated from the mixed
  Cauchy stress $\bsig$, and (d) the effective shear stress $q$. The circles show the
  calculated error for simulations of $n\times n$ resolution where
  $n=900,600,450,360,300,225,200$ and are compared against a reference
  simulation with $n=1800$. The dashed lines show linear fits only using data
  where $h<0.0125$.\label{fig:convergence}}
\end{figure}

\section{Convergence}
\label{sec:convergence}
We measure the convergence of our simulation method by comparing solutions of the flexible rotor simulation of Subsec.~\ref{sub:rotor} as the grid size is reduced, against the solution in a highly refined case. This reference solution plays the role that an exact solution would normally play in such an analysis, but which we do not have given the complexity of the FSI problems that we wish to verify. We have carried out simulations on multiple $n\times n$ grids, for $n=900,600,450,360,300,225,200$. All simulations use $\twid=3\hspc$, so that the width of transition zone via Eq.~\ref{eq:erf2} is fixed in terms of the number of grid spacings. We choose the rotor simulation as a test case and simulate to $t=0.4$, which is long enough for there to be some interaction between the fluid and the solid. At that time, we then compare the fields of each simulation against the corresponding solution of the $1800 \times 1800$ grid, using the discrete $L^2$ norm. For an arbitrary cell-cornered field $f$ this is defined as
\begin{equation}
  ||f^\text{numerical}-f^\text{exact}||_2 = \sqrt{\frac{1}{4^2}\sum_{i=0}^n \sum_{j=0}^n T_iT_j \big| f^{\text{numerical}}_{i,j} - f^{\text{exact}}_{i,j}\big|^2\hspc^2},
\label{eq:L2_norm}
\end{equation}
where \smash{$T_k=\frac{1}{2}$} if $k=0$ or $i=n$ and $T_i=1$ otherwise, so
that the sum calculates the trapezoidal rule. By interpreting the $|\cdot|^2$
operator appropriately, Eq.~\ref{eq:L2_norm} can be applied to scalars,
vectors, or tensors. Since the resolution of each smaller grid is an exact
multiple of the reference grid, each grid point $(i,j)$ on a smaller grid
exactly coincides with a grid point $(i',j')$ on the $1800\times 1800$ grid,
and hence we take $f^\text{exact}_{i,j}=f_{i',j'}$. For an arbitrary
cell-centered field $f$ the same definition
is used but replacing $i,j$ with $[i,j]$. In this case, the smaller grid points may not coincide with the $1800\times1800$
grid points, in which case we calculate $f^\text{exact}$ using
bilinear interpolation.  Figure~\ref{fig:convergence} displays the convergence properties of four key fields. The velocity convergence
plot in Fig.~\ref{fig:convergence}(a) compares the velocity fields in the
entire domain to that of the reference solution. For the convergence plots of
$\bxi$ in Fig.~\ref{fig:convergence}(b), the sum in Eq.~\ref{eq:L2_norm}
is only evaluated at grid points where $\phi<0$ in both the smaller grid and the
reference grid. The figure indicates that all fields are converging with at
least a linear rate, as we expect for the chosen stencils.

Figures~\ref{fig:convergence}(c) and \ref{fig:convergence}(d) show the
convergence of the in-plane pressure $p$ and the in-plane effective shear stress $q=|\sigma_1-\sigma_3|/2$ for $\sigma_1$ and $\sigma_3$ the maximum and minimum principal stresses. The norms are split into
contributions from solid and fluid phases, determined by the sign of $\phi$ on
the coarse grid. In both phases, the stress converges at a similar rate to the
velocity and reference map. For the pressure, the fluid phase has the larger
errors, due to the high value of $\lambda$. However, for the effective shear stress, the solid phase has larger errors, suggesting that errors in solid shear stress are more significant than errors in fluid viscous stress.

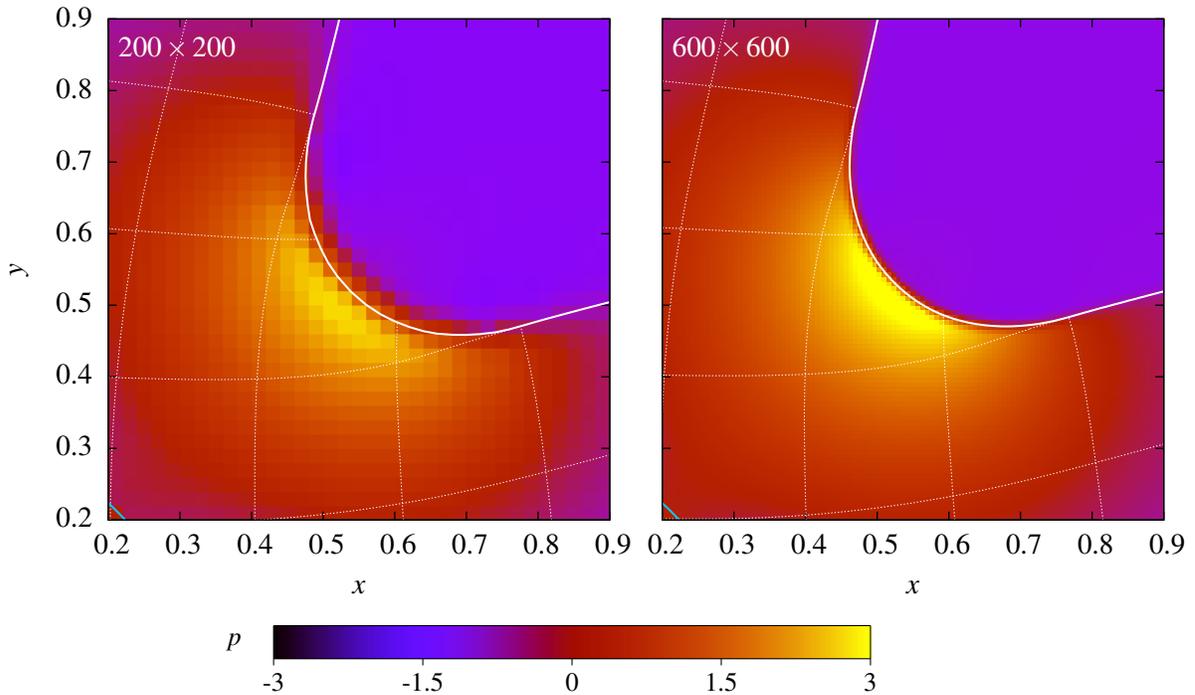
\begin{figure}
  \begin{center}
    \small
    \include{zoom_p}
    \setlength{\unitlength}{0.0125bp}
    \begin{picture}(21100,2000)(0,0)
      \footnotesize
      \put(1600,2000){\includegraphics[width=226pt,height=12.5pt]{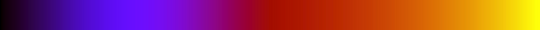}}
      \put(1600,2000){\line(1,0){18000}}
      \put(1600,1800){\line(0,1){1200}}
      \put(1600,3000){\line(1,0){18000}}
      \put(19600,1800){\line(0,1){1200}}
      \put(400,2500){\makebox(0,0)[c]{$p$}}
      \put(6100,1800){\line(0,1){200}}
      \put(10600,1800){\line(0,1){200}}
      \put(15100,1800){\line(0,1){200}}
      \put(1600,1300){\makebox(0,0)[c]{-3}}
      \put(6100,1300){\makebox(0,0)[c]{-1.5}}
      \put(10600,1300){\makebox(0,0)[c]{0}}
      \put(15100,1300){\makebox(0,0)[c]{1.5}}
      \put(19600,1300){\makebox(0,0)[c]{3}}
    \end{picture}
    \vspace{3mm}
  \end{center}
  \caption{Comparison of the pressure field in two simulations of the flexible
  rotor using different grid resolutions. In the left panel, each square that
  is visible corresponds to a grid cell of width $h$, colored according to the
  pressure stored at the cell-centered grid point within the cell. The solid
  white line shows the boundary of the rod, given by the zero contour of the
  level set function. The thin dashed white lines are the contours of the
  reference map $\bxi$. Part of the pivot, shown in cyan, it just visible in
  the bottom left corner.\label{fig:120_1800}}
\end{figure}

It is interesting to zoom into a region containing the interface and compare simulations with a coarse grid to a more refined one. Figure~\ref{fig:120_1800} shows a comparison of the pressure fields at $t=4$ in the upper right concave part of the rotor with the $200 \times 200$ grid and the $600 \times 600$ grid. As the grid size decreases, and likewise the transition zone width, the interfacial behavior approaches that of a sharp interface; to wit, the pressure field develops a rapid variation across the interface that approaches a discontinuity. Analytically, the stress component representing tension in the direction tangent to the interface need not be continuous across the interface, which is why the pressure field should adopt this feature in the sharp limit. 

\begin{figure}
  \begin{center}
    \footnotesize
    \include{extrap1_s11s}
    \setlength{\unitlength}{0.0125bp}
    \begin{picture}(21100,2000)(0,0)
      \footnotesize
      \put(1600,2000){\includegraphics[width=226pt,height=12.5pt]{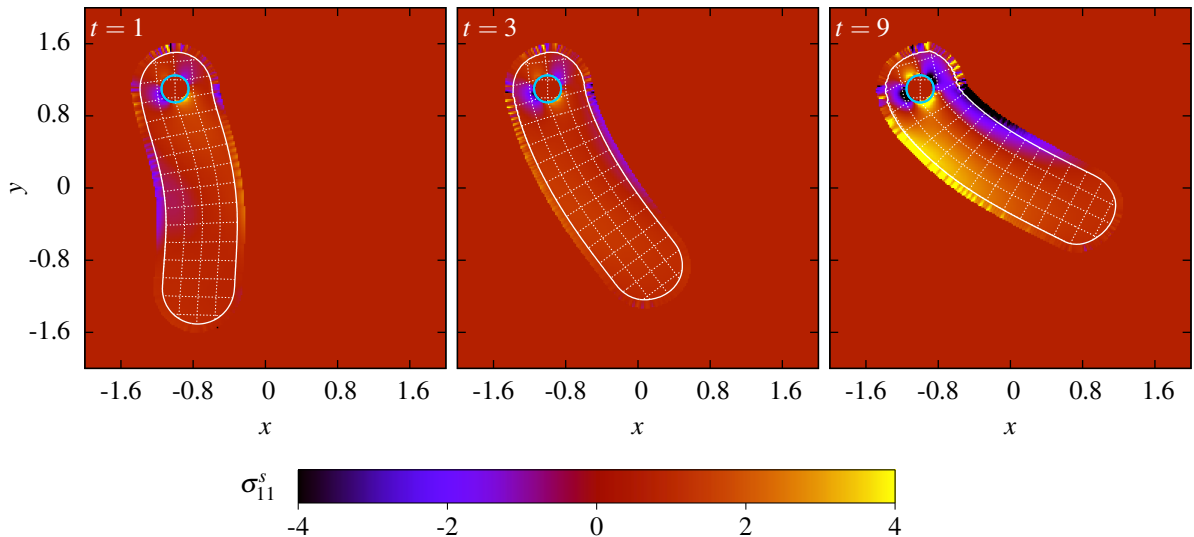}}
      \put(1600,2000){\line(1,0){18000}}
      \put(1600,1800){\line(0,1){1200}}
      \put(1600,3000){\line(1,0){18000}}
      \put(19600,1800){\line(0,1){1200}}
      \put(400,2500){\makebox(0,0)[c]{$\sigma^s_{11}$}}
      \put(6100,1800){\line(0,1){200}}
      \put(10600,1800){\line(0,1){200}}
      \put(15100,1800){\line(0,1){200}}
      \put(1600,1300){\makebox(0,0)[c]{-4}}
      \put(6100,1300){\makebox(0,0)[c]{-2}}
      \put(10600,1300){\makebox(0,0)[c]{0}}
      \put(15100,1300){\makebox(0,0)[c]{2}}
    \put(19600,1300){\makebox(0,0)[c]{4}}
  \end{picture}
  \end{center}
  \vspace{3mm}
  \caption{Three snapshots of the solid stress component $\sigma_{11}^s$ in the
  solid and the extrapolated region, for the anchored rod simulation shown in
  Fig.~\ref{fig:stick_result_2} but where the extrapolation procedure and level
  set motion routine of previous work~\cite{kamrinjmps} are used instead. The solid
  white line shows the boundary of the rod, given by the zero contour of the
  level set function $\phi(\bx,t)$. The thin dashed white lines are the
  contours of the reference map $\bxi(\bx,t)$. The cyan circle of radius 0.15
  shows the anchored region where the reference map is fixed and the velocity
  is zero.\label{fig:first_blurred_s11xe}}
\end{figure}

\section{Improved extrapolation procedure}
\label{extrap_algorithm}
With basic demonstrations in hand, we now return to discuss the details of our reference map extrapolation step. Physically consistent extrapolation of $\bxi$ is key to a proper interface representation, as we shall show. This section highlights two new subroutines we have developed for this purpose. 

The extrapolation algorithm we use is based on that of Aslam~\cite{aslam04}, which takes advantage of the regularized level set $\phi$ already stored on the grid. This routine has the benefit of providing linear extrapolation outside an arbitrary-shaped domain, by extrapolating fields in the outward-normal direction to the interface---the outward normal is obtained easily via $\nor=\nabla\phi$. An immediate requirement of such a routine is that the grid-wise field being extrapolated be smooth near the edge of its known domain. Numerical oscillation error in the known domain, hence, invalidates the extrapolation procedure. This motivates our use of one-sided difference stencils in Sec.~\ref{stencils}, which do not support odd--even oscillation error. Future work will explore higher-order, non-oscillatory stencils to achieve this same aim akin to the usage of staggered grids in computational fluid dynamics. 

While its speed and geometric generality are benefits of the Aslam method, we and others~\cite{rycroft12} have noticed that the hyperbolic nature of the extrapolation routine can cause extrapolated fields to develop striations --- loss of smoothness in the direction parallel to the interface --- even when the known-domain data is smooth. If uncorrected, striations in the extrapolated values of $\bxi$ cause oscillations in the corresponding $\bsig^s$, which induce an artificial wrinkling motion that becomes apparent in the $\phi=0$ level set (see Figure \ref{fig:first_blurred_s11xe}). This phenomenon destabilizes the routine if the wrinkle curvature grows to a level that competes with the grid spacing.

Another important consideration is to ensure agreement between $\bxi$ and the level set field $\phi$ with regard to where the interface lies. We remind that the reference map indicates where material at a current point originated from. If we apply the current reference map field to the initial definition of the level set field, the result should vanish where the current interface lies. To be specific, if we define $\psi$ by $\psi(\bx,t) = \phi(\bxi(\bx,t),t=0)$
then a consistency constraint is
\begin{equation}
  \label{pinning}
  \psi(\bx,t) =0 \iff \phi(\bx,t) =0.
\end{equation}
We note that satisfaction of the above constraint is equivalent to ensuring that all material points initially within the solid remain within the solid.

\begin{figure}
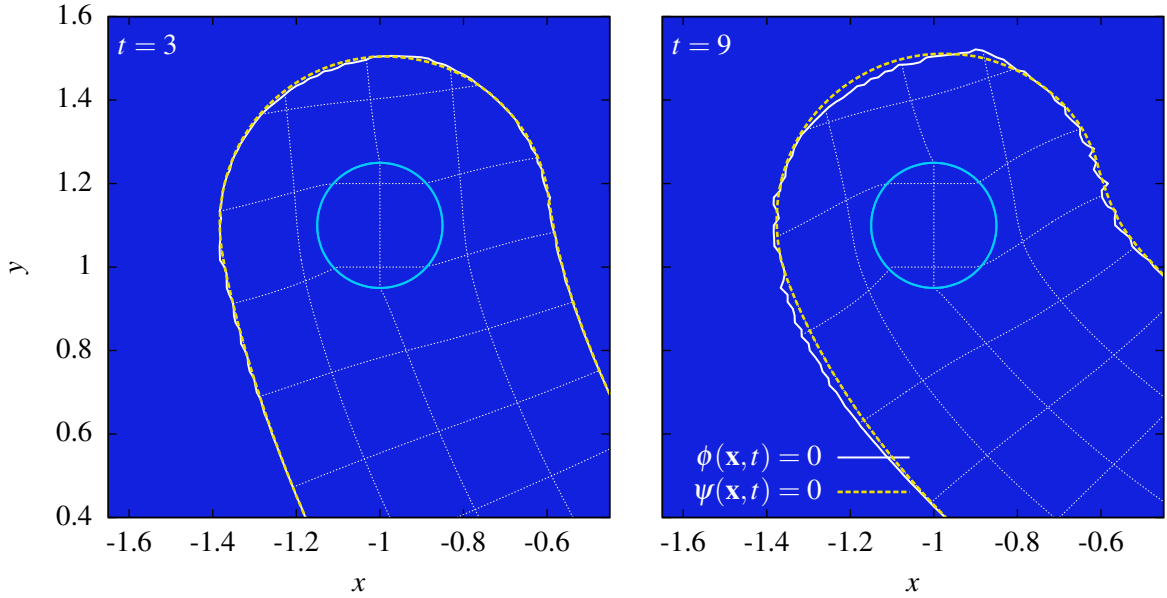

  \begin{center}
    \small
    \include{psi_phi}
  \end{center}
  \caption{Zoomed-in plots of the last two panels of
  Fig.~\ref{fig:first_blurred_s11xe}, displaying drift between the interface as
  described by the level set $\phi(\bx,t)$ and as described by the reference
  map, via $\psi(\bx,t) = \phi(\bxi(\bx,t),0)$. The thin dashed white lines are
  the contours of the reference map $\bxi(\bx,t)$. The cyan circle of radius
  0.15 shows the anchored region where the reference map is fixed and the
  velocity is zero.\label{fig:drift_phi_psi}}
\end{figure}

Because the values of $\bxi$ and $\phi$ are obtained from different numerical routines, discretization error can cause the above constraint to be violated over time.  Recall that $\bxi$ is generated on points with $\phi>0$ solely through linear extrapolation from the adjacent $\phi<0$ domain, and hence the values of $\psi$ near the zero-contour of $\phi$ are not more than first-order accurate. The comparison between them in Fig.~\ref{fig:drift_phi_psi} is indicative that a drift between $\psi$ and $\phi$ grows after the onset of the previously described wrinkling artifact.  This in turn causes persistence of the wrinkled shape, because some material that began on the solid side of the $\phi=0$ interface is falsely assigned as fluid, which permits the wrinkles to sustain themselves over time without the elastic response correcting their shape.  This calls for a routine to ensure the \emph{physical requirement} that the zero contours of $\phi$ and $\psi$ satisfy Eq \ref{pinning}, i.e. remain ``pinned'', to ensure that solid stays solid and fluid stays fluid. One could argue at this moment that $\phi$ is a redundant field, in that $\psi$ could operate just as well on its own at discerning phases. However we recall that $\phi$ also gives a measure of distance, as its gradient always has magnitude 1. This feature is exploited in the Aslam extrapolation scheme and is needed for the solid--solid contact algorithm of Sec.~\ref{extrap_algorithm}.  

We now describe two mid-step subroutines, which are applied after an initial Aslam extrapolation of $\bxi$ is conducted, and successfully resolve the interfacial wrinkling, phase exchange error, and instabilities relating to these two phenomena. The need to appropriately pin the reference map extrapolation field and the level set field was pointed out previously~\cite{kamrinjmps} but a general and accurate pinning technique was not employed in that work. Artificial surface tension, projection of the extrapolated reference map field, and solid damping were used to reduce the appearance of these problems in that work.

The first step in our procedure is to remove artificial striations in the extrapolated values, see Algorithm 2. It is a general algorithm that can be applied to to reduce fluctuations in an arbitrary extrapolated field $f$. In the FSI routine, it is applied separately to $\xi_x$ and $\xi_y$. The fields on the solid side of the interface are never affected; only the extrapolated values within the fluid domain are adjusted. The Algorithm consecutively applies a diffusion stencil to improve the extrapolated values. The edge of the solid domain is sampled by the diffusion stencil and hence the resulting extrapolation is both smooth within the fluid domain and continuously extends data from the solid domain. 

\begin{algorithm}[t]
\textit{Given:} \ {A field $f_{i,j}$ defined where $\phi_{i,j}<0$, and an extrapolation of $f$ into the region $\phi_{i,j}>0$}
\\
\textit{Compute:} \ {A smoother extrapolation of $f$ where $\phi_{i,j}>0$}\\

\begin{enumerate}
  \item Define a starting field $f^0_{i,j}=f_{i,j}$ at all points in the region
    where $\phi_{i,j}>0$ and all four orthogonally adjacent neighbors are part
    of the extrapolated region.
  \item For $k=1, 2, \ldots, 5$ calculate $f^{k+1}_{i,j} =
    0.05(f^k_{i,j-1}+f^k_{i-1,j}+f^k_{i+1,j}+f^k_{i,j+1}-4f^k_{i+1,j})$.
  \item Return $f^{5}_{i,j}$ as the smoothed extrapolation.
\end{enumerate}
\caption{Extrapolation smoothing subroutine.\label{alg:diffuse}}
\end{algorithm}
    
\begin{figure}
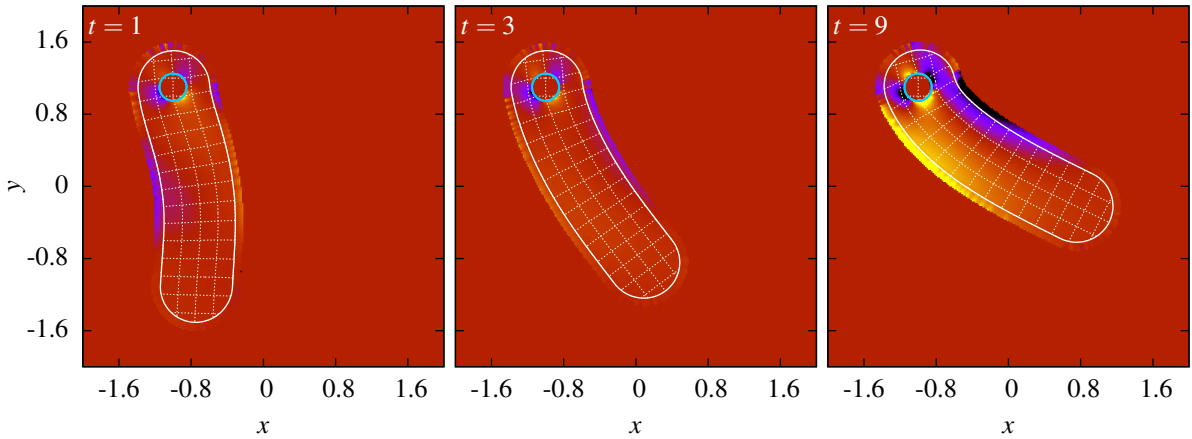

  \begin{center}
    \footnotesize
    \include{extrap2_s11s}
  \end{center}
  \caption{Three snapshots of the solid stress component $\sigma_{11}^s$ in the
  solid and the extrapolated region, for the anchored rod simulation of
  Fig.~\ref{fig:stick_result_2} that uses the new extrapolation procedure. The
  color gradient is the same as in Fig.~\ref{fig:first_blurred_s11xe}. The
  solid white line shows the boundary of the rod, given by the zero contour of
  the level set function $\phi(\bx,t)$. The thin dashed white lines are the
  contours of the reference map $\bxi(\bx,t)$. The cyan circle of radius 0.15
  shows the anchored region where the reference map is fixed and the velocity
  is zero.\label{fig:extrap_fixed}}
\end{figure}

Upon completion of Algorithm 2, the second mid-step routine starts by updating the level set values $\phi(\bx,t)$ in the narrow band to be equal to $\psi(\bx,t)$. This ensures that the zero contour of the level set is consistent with the reference map, but the new values of $\phi(\bx,t)$ may not satisfy the signed distance function property, $|\nabla \phi|=1$. We therefore make use of the reinitialization routine described in ~\cite{rycroft12}, to rebuild the narrow-banded level set function and recover this property. Figure~\ref{fig:extrap_fixed} displays the solid stress from the reference map and its extrapolation in the same test geometry but using our new two-part extrapolation-pinning routine. The formerly observed striated solid stress, interfacial wrinkles, and drift between fields have been eliminated.

\section{Simulation of contact between two solids immersed in a fluid}\label{FSSI}
We now consider a second solid phase and seek to simulate a fully coupled fluid--solid--solid interaction; solids interacting through contact while submerged in a fluid. We describe the procedure for two solid phases, but the procedure could be generalized to more solids. We use superscripts $(1)$ and $(2)$ to denote each solid phase. There are now two level sets, \smash{$\phi^{(1)}_{i,j}$} and \smash{$\phi^{(2)}_{i,j}$}, to describe each solid's boundary and measure distance to it. Each level set has a corresponding transition field, $\Trans^{(1)}$ and $\Trans^{(2)}$, using the same $n_b$ for both. We maintain two reference map fields, \smash{${\bxi}^{(1)}_{i,j}$} and \smash{${\bxi}^{(2)}_{i,j}$}, and corresponding solid stresses, \smash{${\bsig}^{s (1)}_{[i,j]}$} and \smash{${\bsig}^{s (2)}_{[i,j]}$}, within the extended domain of each associated solid. The velocity field, as before, is a single unique field for the whole computational domain, independent of phase.  
The main procedure, Algorithm 1, requires very little change to represent interacting submerged solids. All steps corresponding to the solid are now simultaneously performed on both sets of solid fields. The major changes we must prescribe are to adjust the mixing rule in Step 8 to correctly cross-fade between the two solids and fluid, and to add an extra routine to correctly set contact conditions. The latter consideration ultimately arises as an adjustment in Step~6.

\subsection{Mixing formulation}
The transition function of solid phase $(i)$ indicates at any location the fraction of the material behavior attributable to that phase.  Therefore, in the presence of two solid phases and fluid, we arrive at the below three-way mixing protocol,
\begin{equation}
\bsig_{[i,j]} = \Trans^{(1)}_{[i,j]}\bsig^{s(1)}_{[i,j]}+ \Trans^{(2)}_{[i,j]} \bsig^{s(2)}_{[i,j]}+\left(1-\Trans^{(1)}_{[i,j]}-\Trans^{(2)}_{[i,j]}\right) \bsig^{f}_{[i,j]}
\label{eq:fssi_mix}
\end{equation}
on each cell-centered grid point; the quantity of each solid follows the previous behavior, and the remainder is filled with fluid. The coefficient of the third term remains non-negative so long as the solid phases  do not penetrate. The same approach is applied in defining the density. 

\subsection{Setting contact conditions}
The contact algorithm focuses on the case of frictionless contact, but it is sufficiently general that other contact conditions could be implemented. In the blurred interface routine, we define contact between two objects whenever there is an overlap between the transition zone of one object and the interface of the other. We define penetration to occur if the zero contours of each level set field pass through one another. Hence, a non-penetrating contact routine permits the zero level sets of the two bodies to be separated by less than half the transition zone width -- the `start' of blurred contact -- and rapidly penalizes any closer approach of the two solids.  When the solids are farther apart than half the transition zone width, the routine should have no effect. As the grid size shrinks, this description approaches the standard definition of non-penetrating contact between sharp interfaces.

The key idea is to exploit the quantity defined as the difference between the two solid level set functions, 
\begin{equation}
  \phi_{12}= \frac{\phi^{(1)}- \phi^{(2)}}{2}.
\label{eq:dphi}
\end{equation} 
Note that $\phi_{12}=0$ implies a point in space equidistant between the two solid boundaries, and we refer to the set of all such points as the mid-surface between the two bodies. At a point on the boundary of phase 1, the value of $2|\phi_{12}|$ indicates the distance to the nearest point on the boundary of phase 2, and {\it vice versa}. This is true regardless of the geometry of the two solids. This property lets us use $\phi_{12}$ to construct a short-range separation force that captures our desired contact condition. We define a compactly supported influence function as
\begin{equation}
  \delta_s(x)= \left\{
  \begin{array}{ll}
    \dfrac{1+\cos\frac{\pi x}{\twid}}{2\twid} & \qquad \text{if $|x|<\twid$,} \\
    0 & \qquad \text{if $|x| \ge \twid$},
  \end{array}
  \right.
\end{equation}
so that $\delta_s(x)$ is the derivative of $H_s(x)$ as defined in
Eq.~\ref{eq:heavi_smooth}. The above is used to define a mutually repulsive
force field centered on the mid-surface, which acts only on solid points that
intersect its influence. Mathematically, this is achieved by defining the
separation function
\begin{equation}
  \Sep_{i,j} = \krep \delta_s(\phi_{12 i,j}),
\end{equation}
from which the body force field is defined as
\begin{equation}
  \vec{f}_{i,j}=
  \left\{
  \begin{array}{ll}
    \Sep_{i,j} \nor_{12 i,j} & \qquad \text{if $\phi^{(1)}_{i,j}<0$ or $\phi^{(2)}_{i,j}<0$,} \\
    \vec{0} & \qquad \text{otherwise,} \\
  \end{array}
  \right.
\end{equation}
which is used within Eq.~\ref{eq:velocity_update}, where \smash{$\nor_{12
i,j}=\pm \frac{\nabla \phi_{12 i,j}}{|\nabla \phi_{12 i,j}|}$} is the vector normal
to the contours of $\phi_{12}$, pointing away from the mid-surface. The
prefactor $\krep$ must be chosen large enough to successfully repel a crossing
of the interface-centers, but must be small enough that the stable
time-increment of such an explicit contact routine also decreases with $\krep$.
In the following examples we choose \smash{$\krep=\frac{1}{20}$}.

\subsection{Computational results}
\begin{figure}
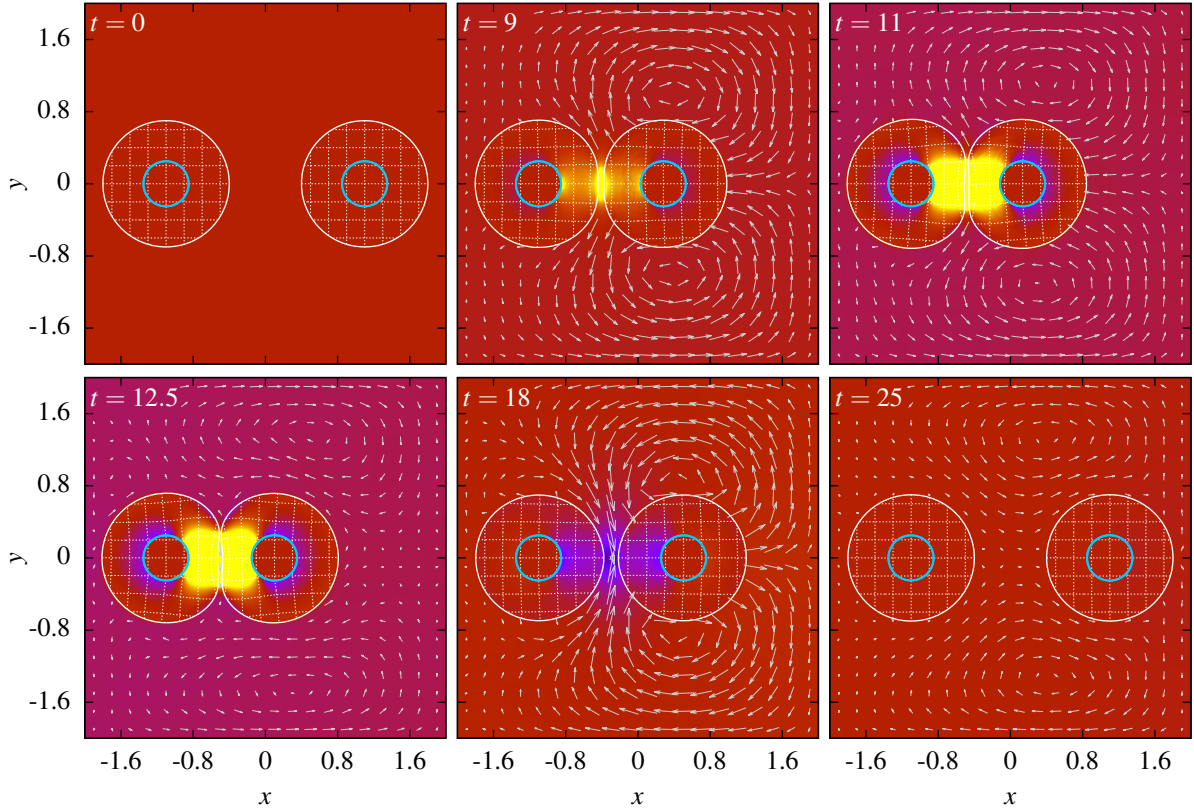

  \begin{center}
    \footnotesize
    \include{contact_p}
  \end{center}
  \caption{Six snapshots of the pressure field (colors) and fluid velocity
  (arrows) for a simulation where a moving disk comes into contact with an
  anchored disk. The lengths of the arrows are proportional to
  \smash{$\sqrt{|\bv|}$}. The colors for the in-plane pressure use
  the same key as in Fig.~\ref{fig:stick_result_1}. The solid white lines show
  the boundaries of the two disks, given by the zero contours of the level
  set functions. The thin dashed white lines are the contours of $\bxi(\bx,t)$ defined within the two disks. In each snapshot, the left
  cyan circle of radius 0.25 is anchored, and the right cyan circle of radius
  0.25 is moving.\label{fig:ball_ball_good}}
\end{figure}

We first consider two submerged, colliding neo-Hookean disks of radius 0.7 that are initially centered at $(x,y)=(\pm 1.1,0)$, using a $256\times 256$ simulation grid and simulating over the interval $0\le t \le 25$. Each disk has a circular anchoring region of radius 0.25 at its center. The left disk's anchoring region is permanently fixed, while the right disk's anchoring region has a time-varying horizontal position
\begin{equation}
x_d(t) = 1.1 - 0.5\left(1-\cos \frac{2\pi t}{25} \right)
\end{equation}
and horizontal velocity
\begin{equation}
v_d(t) = - \frac{\pi}{25} \sin \frac{2\pi t}{25}.
\end{equation}
Figure~\ref{fig:ball_ball_good} shows six snapshots of pressure during the simulation. By $t=9$, the pressure starts to build up between the two disks as they come into contact. By the time the moving disk reaches its leftmost position at $t=12.5$ when $x_d(t)=0.1$, the pressure has increased further. At $t=18$, the moving disk has separated from the anchored disk, and a small region of negative pressure is created between the two disks as fluid. At $t=25$, the moving disk comes to rest at its original position, and returns to its original undeformed circular shape.

\begin{figure}
  \begin{center}
    \footnotesize
    \include{contactz_ffu}
    \setlength{\unitlength}{0.0125bp}
    \begin{picture}(21100,2000)(0,0)
      \footnotesize
      \put(1600,2000){\includegraphics[width=226pt,height=12.5pt]{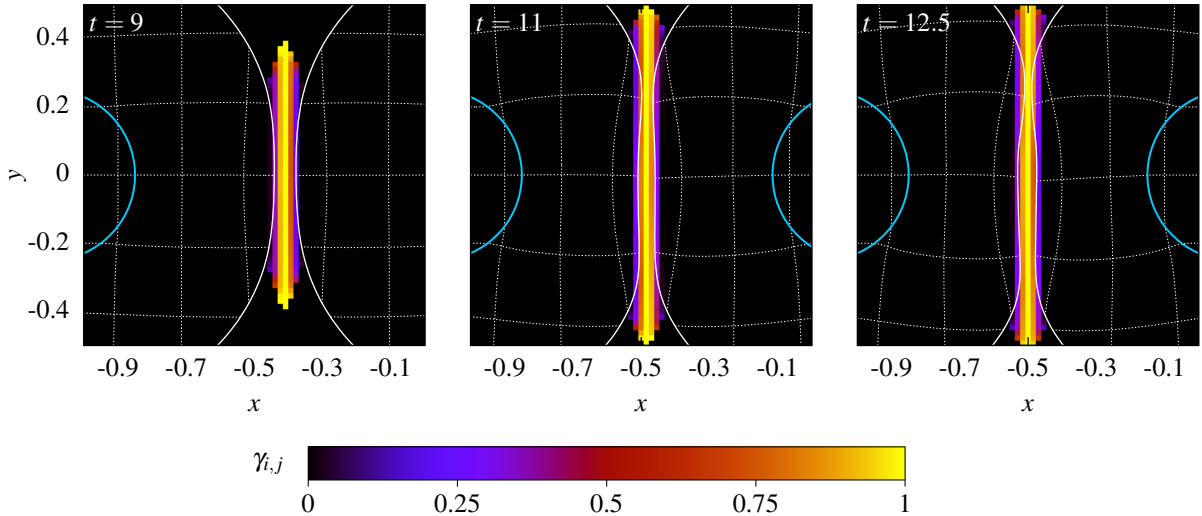}}
      \put(1600,2000){\line(1,0){18000}}
      \put(1600,1800){\line(0,1){1200}}
      \put(1600,3000){\line(1,0){18000}}
      \put(19600,1800){\line(0,1){1200}}
      \put(400,2500){\makebox(0,0)[c]{$\Sep_{i,j}$}}
      \put(6100,1800){\line(0,1){200}}
      \put(10600,1800){\line(0,1){200}}
      \put(15100,1800){\line(0,1){200}}
      \put(1600,1300){\makebox(0,0)[c]{0}}
      \put(6100,1300){\makebox(0,0)[c]{0.25}}
      \put(10600,1300){\makebox(0,0)[c]{0.5}}
      \put(15100,1300){\makebox(0,0)[c]{0.75}}
      \put(19600,1300){\makebox(0,0)[c]{1}}
    \end{picture}
  \end{center}
  \vspace{2mm}
  \caption{Three snapshots showing a zoomed-in region of the separation
  function $\Sep_{i,j}$ in the simulation of two submerged, colliding disks.
  The function is only defined in the region where the narrow bands for the two
  level sets overlap, and is plotted as zero outside this region. The solid
  white lines show the boundaries of the two disks, given by the zero contours
  of the level set functions. The thin dashed white lines are the contours of
  the reference map $\bxi(\bx,t)$ defined within the two disks. In each
  snapshot, the left cyan circle of radius 0.25 is anchored, and the right cyan
  circle of radius 0.25 is moving.\label{fig:red_line}}
\end{figure}

Figure~\ref{fig:red_line} displays the $\Sep_{i,j}$ function at three time points, and shows how the solids do not feel the interaction until the moment when the narrow $\delta_s$ function enters into the solids. Within the simulation, the function is only defined in the region where the two narrow bands of the level set overlap. The repulsion force succeeds in separating the two interface-centers but maintains contact in the sense defined above. Some small asymmetry is visible, which is expected due to the asymmetry in the stencils described in Fig.~\ref{fig:rmap_grad2}; this effect diminishes with grid spacing. As is evident in Fig.~\ref{fig:ball_ball_good} using this technique we are now capable of getting very large deformations in the solids through contact, without any problem of penetration or sticking of one solid onto the other.

The second example that we consider is a disk bouncing on an anchored rubber bar, all while submerged in fluid. The disk has radius 0.4 and is initially centered at $(x,y)=(1,0)$. The bar comprises of the rectangle $-1<x<1, -0.9<y<-0.1$ along with semi-cicular end caps, and is anchored in a circle of radius $0.2$ at $(x,y)=(-1,0.5)$. For this example, the density of the disk is 20, the disk's initial downward velocity is $-1$, the density of the rod is 6, the gravity is 0.03, the fluid viscosity is $\eta=0.04$. All other parameters are unchanged from the previous results. A grid of $600\times 600$ is used and the system is simulated over the range $0\le t\le 25$.

\begin{figure}
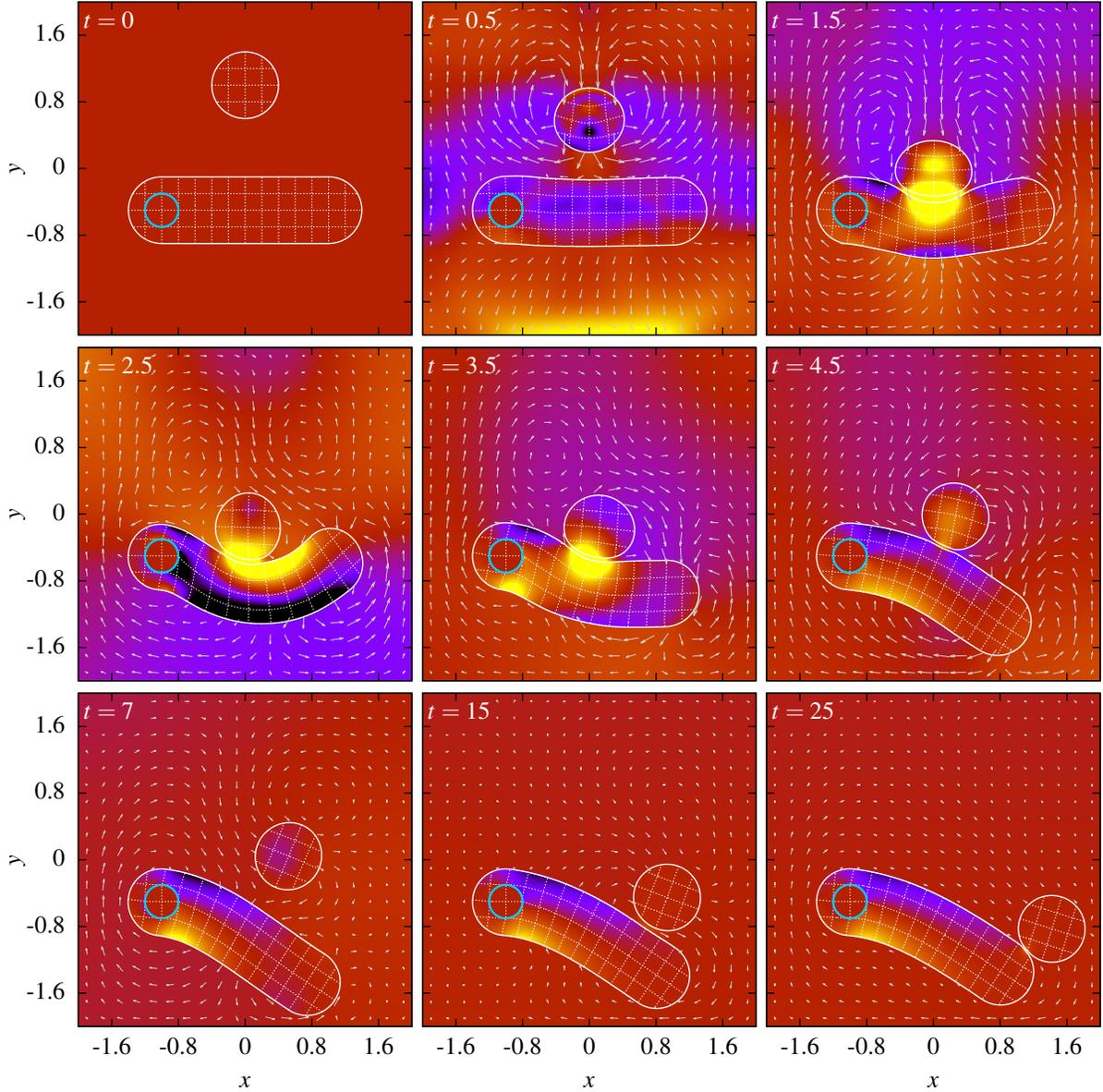

  \begin{center}
    \footnotesize
    \include{paddle_p}
  \end{center}
  \caption{Nine snapshots of the pressure field (colors) and fluid velocity
  (arrows) for the simulation of ball fired into an anchored flexible rod. The
  length of the arrows are proportional to \smash{$\sqrt{|\bv|}$}, where a
  nonlinear scaling is used to the large variations in the size of the
  velocity. The colors for the pressure use the same key as in
  Fig.~\ref{fig:120_1800}. The solid white lines shows the boundary of the rod
  and the circle, given by the zero contour of the level set functions. The
  thin dashed white lines are the contours of $\bxi(\bx,t)$
  defined within each of the two objects. The cyan circle of radius 0.2 shows
  the anchored region of the rod where the reference map is fixed and the
  velocity is zero.\label{fig:paddle_p}}
\end{figure}

Figure~\ref{fig:paddle_p} shows nine snapshots of the in-plane pressure field during this simulation. The disk first reaches the rod at approximately $t=1.5$, and a large region of positive pressure is visible between the two. By $t=2.5$ the force exterted by the heavy disk deforms the bar into a U-shape, with a large tension visible on the bottom side of the bar. By $t=4.5$ the right end of the bar has moved downward, and this motion pushes the disk upward so that it undergoes a bounce. By $t=7$ the disk is fully separated from the bar. The disk then sinks and comes into contact with the bar at $t=15$, before slowly sliding down the bar.

This approach allows us to model frictionless non-sticking contact that avoids cross-penetration of the solid phases. We could imagine adherence conditions, friction or other contact laws through new definitions of a possibly evolutionary influence function centered at the mid-surface. Other Eulerian contact approaches also exist; we can for example measure the amount of overlap a step would cause and if any then apply a correction based on the intersecting volume~\cite{levin11} and a minimization problem finding the optimal velocity field which solves the equilibrium condition while minimizing overlap. The ease of contact detection and penalization in Eulerian frame, through use of pre-existing interfacial distance functions, is indeed an advantage of the Eulerian approach. 

\section{Conclusion}
This work has described a blurred-interface finite-difference method for fluid--structure interaction on a single Eulerian grid. The method is notable for its simplicity and speed. Our explicit algorithm invokes a computation of fluid and solid stress, which are then mixed in accordance with a transition function, as are the fluid and solid density. The solid stress and density are computed in Eulerian-frame with the aid of the reference map field, which is stored and updated throughout. An important pair of subroutines  is also presented, to improve the representation of the reference map and level set fields near the interface. Our method is shown to converge and as grid-size decreases we recover results that properly display the signatures of a sharp interface. The framework we create extends to the case of solid--solid contact as well and we have given two examples of submerged contact for the case of non-penetrating, frictionless, non-sticking behavior. Here, the key idea is to take advantage of the distance-function property of the level set fields about each solid to produce a short-range separation force that acts only within the solid interfacial regions when they are close enough to each other.

There are a number of important directions to consider from this point. From a numerics standpoint, we can go to higher convergence rates by moving to a higher-order set of finite-difference stencils, though we emphasize the importance of removing oscillatory numerical errors, as we discussed in Sec.~\ref{extrap_algorithm}, which can arise in higher-order stencils. It is also important to port our scheme into 3D, which should require very little algorithmic adjustment. We can also implement incompressibility constraints by adapting the projection method~\cite{chorin68}.  It would be worthwhile to parallelize the scheme beyond multi-threading to take advantage of a distributed-memory architecture.  While we have focused on Cartesian grids for ease, this is not a necessity, and it would be useful to consider general meshes that enable local refinement. 

There are also several next steps to be taken from the applications standpoint. We are actively pursuing sub-modeling capabilities within this approach, which model the domain using two different grids of different refinement levels, and enable us to refine a local zone without the need to decrease the time-step in the coarse grid zone. This would enable local refinement near a rough fluid--solid interface and could be applied to multi-scale problems such as nuclear fuel-rod fretting, which involve interactions between cooling fluid and roughened solid parts.   To extend the solid contact routine to the case of many submerged solid phases could have considerable usage in modeling dense suspensions of soft particles.  To be able to model contact-induced finite-deformations of particles is crucial when the compliance of particles is too high to admit simple contact-force laws.  Biological applications, particular at the cellular level, could offer another arena in which this approach could be of use, as the solid components are highly deformable and fluid permeates the system.  The Eulerian form could be advantageous in the implementation of growth models \cite{garikipati09},  or simultaneous species diffusion, which is also amenable to Euerlian-frame.

\section{Acknowledgements}
B.~V. and K.~K. acknowledge support from the MIT Department of Mechanical
Engineering. K.~K. acknowledges support from the Consortium for Advanced
Simulation of Lightwater Reactors (CASL) an Energy Innovation Hub for Modeling
and Simulation of Nuclear Reactors under US Department of Energy Contract No.
DE-AC05-00OR22725. C.~H.~R. was supported by the Director, Office of Science,
Computational and Technology Research, U.S.~Department of Energy under contract
number DE-AC02-05CH11231.

\newpage

\bibliography{solids2}

\end{document}